\documentclass[journal]{IEEEtran}
\usepackage{mathtools}
\usepackage[margin=1in]{geometry}

\usepackage{comment}

\usepackage{epsfig,graphicx,subcaption,color,url} 

\usepackage{algorithm}
\usepackage{graphics}
\usepackage{epsfig}
\usepackage{wrapfig}
\usepackage{cite}
\usepackage{algpascal}
\usepackage{algc}
\usepackage{algcompatible}
\usepackage{multicol}
\usepackage{algpseudocode}
\usepackage{subcaption}
\usepackage{wrapfig}
\usepackage{xurl}
\usepackage{hyperref}
\usepackage{orcidlink}
\usepackage{soul}
\setstcolor{red}
\usepackage{lettrine}
\DeclareMathAlphabet{\pazocal}{OMS}{zplm}{m}{n}
\usepackage{comment}
\usepackage[normalem]{ulem}

\usepackage{amssymb,amsmath,algorithm}
\usepackage{amsfonts,amsthm, bm, bbm,dsfont}

\usepackage{xcolor}

\ifCLASSINFOpdf
\else
\fi
\begin{document}
%
\title{Bootstrap-Based Receiver Synchronization and System Discovery in B2X: An Extension of ATSC 3.0}
\author{\IEEEauthorblockN{Raj Kumar Thenua\orcidlink{0000-0001-6011-4390}\IEEEauthorrefmark{1}, 
Essam Sourour\orcidlink{0000-0002-4473-7026}\IEEEauthorrefmark{2}, David Starks\orcidlink{0009-0002-5854-1283}\IEEEauthorrefmark{2}, Rashmi Kamran\orcidlink{0000-0003-0679-7631}\IEEEauthorrefmark{3}, \\
Michael Simon\orcidlink{0000-0003-0881-8811}\IEEEauthorrefmark{4}, Kumar Appaiah\orcidlink{0000-0002-3149-4416}\IEEEauthorrefmark{1}}\\
\IEEEauthorblockA{Department of Electrical Engineering,
Indian Institute of Technology Bombay, Mumbai, India\IEEEauthorrefmark{1}\\
\IEEEauthorblockA{AdCore Local, LLC., Norcross, GA 30092, USA\IEEEauthorrefmark{2},
\IEEEauthorblockA{EdgeBeam Wireless, Boston, MA 02210, USA\IEEEauthorrefmark{3}\\ 
\IEEEauthorblockA{Sinclair, Inc., Cockeysville, MD 21030, USA\IEEEauthorrefmark{4}}}
\\Email: 30006253@iitb.ac.in\IEEEauthorrefmark{1}, sourour@adcorelocal.com\IEEEauthorrefmark{2}, dstarks@adcorelocal.com\IEEEauthorrefmark{2}, rkamran-c@edgebeamwireless.com\IEEEauthorrefmark{3},
msimon@sbgtv.com\IEEEauthorrefmark{4}, akumar@ee.iitb.ac.in\IEEEauthorrefmark{1}
\vspace{-1cm}
}}}
\maketitle

\begin{abstract}
Addressing the increasing and diversified demands of multicast and broadcast services requires highly efficient multicast and broadcast technologies. Broadcast networks, such as the Advanced Television Systems Committee (ATSC) 3.0, are inherently designed to support these services and continue to evolve to meet growing performance and scalability requirements. At the same time, smartphones are increasingly used for video streaming and other high-volume services, placing growing pressure on mobile network capacity. Interworking between broadcast and mobile networks is therefore an important enabler for efficient and seamless service delivery. In this context, Broadcast-to-Everything (B2X) extends ATSC 3.0 to support enhanced interoperability with Third Generation Partnership Project (3GPP) mobile systems while maintaining 
low cross-correlation with ATSC 3.0 bootstrap signals, supporting reliable system identification in scenarios where multiple waveforms may be present. Bootstrap signaling, which enables initial signal detection and synchronization, is a key feature of ATSC-based waveform discovery and synchronization, and B2X further extends this capability 
through a scalable bootstrap framework supporting a range of bandwidth configurations.
This paper investigates system discovery through bootstrap signal detection at the B2X receiver and presents key design-related findings, including parameter selection and cross-testing with ATSC 3.0. We present extensive simulations of the receiver performance under diverse propagation and mobility conditions, ranging from stationary to high-speed scenarios. The results demonstrate the robustness of the B2X bootstrap signaling design across a broad range of channel conditions relevant to multicast and broadcast operation.
\end{abstract}


%
\IEEEpeerreviewmaketitle
\section{Introduction}
\lettrine{I}{n} recent years, high‑resolution video consumption, live event streaming, and over-the-air (OTA) software and firmware updates have experienced rapid growth. This trend is expected to continue, with the projected number of users in the over-the-top (OTT) video market expected to reach 4.9 billion by 2029~\cite{muvi}. 
The use of mobile phones for accessing such services has also increased rapidly, with future generations of consumers (nearly 90\% of Gen-Z viewers) expected to use smartphones to watch videos and access streaming services\cite{streamable}.
Among mobile network applications, video accounts for 82\% of total consumer traffic \cite{ericcson}. 
\par Integration of technologies and multi-platform capabilities are therefore key drivers in meeting these demands. These demands have placed significant stress on existing mobile network infrastructure. Multicast and broadcast delivery methods offer a more efficient means of distributing identical content to large numbers of users simultaneously. Furthermore, public safety and emergency alert systems require reliable, wide-area dissemination, a role for which broadcast delivery is particularly well suited. As mobile device usage continues to expand, offloading large-scale content delivery from unicast to multicast and broadcast becomes increasingly essential.
\par To support multicast and broadcast services, Third Generation Partnership Project (3GPP) has standardized a framework known as 5G Multicast-Broadcast Services (5G MBS) [3GPP TS 23.247], which enables efficient delivery of identical content to multiple users within mobile networks. In practice, however, network resources must be shared between unicast and multicast/broadcast delivery methods. As a result, integration of mobile networks with broadcast technologies can enhance overall network resilience by enabling access to supplementary delivery infrastructures\cite{kamran2022}.  
\par ATSC 3.0 is an Internet Protocol (IP)-native broadcast system supporting both traditional television and IP-based services, with flexible service delivery and hybrid broadcast–broadband operation \cite{Chernock_2016}. These capabilities provide a foundation for extending broadcast systems toward more scalable and flexible operation.
\subsection{B2X: An extension to ATSC 3.0} A new release of ATSC referred to as Broadcast-to-Everything (B2X), extends ATSC technology towards seamless convergence between broadcast and mobile \cite{ibc}. B2X incorporates modern Forward Error Correction (FEC) schemes derived from Fifth Generation (5G) New Radio (NR) including Low‑Density Parity‑Check (LDPC) and Polar codes, and employs flexible physical-layer resource multiplexing mechanisms to support multiple services and deployment scenarios. Its radio access network (RAN) design principles are intended to facilitate interoperability with cellular network architectures. B2X is suitable for both fixed and portable receivers, particularly in high-density or underserved areas. It introduces a scalable bootstrap design to support a wide range of device capabilities, from low-power devices to automotive receivers. Additionally, B2X provides a unified framework for hybrid receivers to access both broadcast and 3GPP-based services in a complementary manner, enabling efficient broadcast offload and interoperability with emerging 5G/International Mobile Telecommunications (IMT)-2030 systems. { Beyond 5G offload, identified B2X use cases include IoT firmware and software distribution, automotive over-the-air (OTA) updates and HD map distribution, emergency and public safety communications, AI model distribution, AR/VR asset distribution, and UAV/drone communications  \cite{IITB_whitePaper}.}
\par B2X is designed to support operation across a range of spectrum allocations, including traditional UHF broadcast bands as well as IMT bands used in mobile systems. This capability enables flexible deployment across diverse regulatory and service environments, and supports convergence scenarios in which broadcast and mobile delivery mechanisms operate in an integrated manner.   A unique feature of ATSC B2X technology is its in‑frame signaling mechanism based on bootstrap signaling~\cite{He_2016}. The bootstrap signal is transmitted at the beginning of each frame period and supports functions such as initial service discovery, channel estimation, and receiver synchronization. The bootstrap configuration defines parameters including sampling rate, bandwidth, subcarrier spacing, and time‑domain structure. Using this information, the receiver determines the configuration of the waveform that follows, enabling the broadcast spectrum to evolve toward new signal types and services. B2X incorporates specific design changes to the ATSC 3.0 bootstrap to enable scalable operation across a range of bandwidth configurations. 
\par In this paper, we present key B2X signal discovery and receiver design aspects, with particular emphasis on synchronization and system discovery through bootstrap signaling. The primary focus is to examine the robustness of the B2X bootstrap design across diverse usage scenarios and channel conditions. {We limit our performance evaluation to the sub-1-GHz frequency range.}  
\subsection{Literature survey}
This section summarizes prior work related to synchronization and signal detection in ATSC 3.0, 3GPP 5G NR, and B2X systems. Reliable system discovery and synchronization are fundamental requirements for receivers in wireless communication systems, particularly in broadcast deployment scenarios where the receiver must detect and acquire signals without prior coordination.
Receiver synchronization involves both time and frequency acquisition, enabling alignment to the transmitted waveform and subsequent decoding of signaling and data. These functions are typically supported through structured reference or bootstrap signals, whose design directly influences detection performance, robustness to channel impairments, and receiver complexity. The following review focuses on synchronization and detection approaches relevant to receiver-side system discovery.
In 3GPP 5G NR systems, User Equipment (UE) synchronization is achieved through over-the-air reference signals, including the Synchronization Signal Block (SSB), which supports initial cell detection, timing acquisition, and frequency synchronization. Additional reference signals such as demodulation and channel state information are used for channel estimation and tracking during ongoing operation. These mechanisms enable reliable receiver synchronization under a range of deployment conditions.
\par A significant body of prior work has contributed to understanding synchronization, receiver design, and physical-layer operation in ATSC 3.0 since the introduction of the system discovery and signaling framework over a decade ago \cite{standard2016system}. An overview of ATSC 3.0 that includes many of the related works cited is provided in ~\cite{Chernock_2016}. The physical-layer design of ATSC 3.0 supports a range of coverage and service performance requirements for digital broadcast systems, and offers various 
 trade-offs through flexible parameterization and framing structures~\cite{Fay_2016,Earnshaw_2016}. Design choices were made with receiver complexity in mind, and various detection and synchronization schemes have been studied to balance performance and implementation cost~\cite{Kim_2019}. System discovery and signaling transmission using the bootstrap approach are discussed in~\cite{He_2016}. 
These approaches have been shown to be competitive, and in many cases more appropriate for broadcast transmission, when compared to traditional cellular communication–based approaches~\cite{Kwon_2025,Ahn_2023}.
While the physical layer and system discovery and signaling aspects of ATSC-based broadcast systems have been extensively studied, demonstrating adequate performance requires comprehensive simulation and experimental validation across a range of representative channel models and propagation conditions. To this end, substantial effort has been devoted to characterizing the performance of ATSC-based communication systems under diverse channel environments \cite{Kwon_2025,Fay_2016}.
\par Implementation guidelines and baseline channel models for comparison with Digital Video Broadcasting (DVB)-T2 systems are provided in~\cite{etsi2004102}. More recently, standardized 5G NR channel models applicable to broadcast frequency bands have been documented in~\cite{3gpp_tr38901_v18}. In addition, empirical channel characterization in Ultra High Frequency (UHF) Single-Frequency Network (SFN) environments has been reported in~\cite{ahn2022characterization}. Given the strong alignment among standards, applications, and achievable data rates, the channel models and comparative baselines described in these references are adopted in this work to evaluate the performance of the B2X receiver approaches.
\par ATSC B2X has been introduced conceptually in a limited number of publications \cite{ibc,R4Ref1,R4Ref2}. {   There is a specialist group within the ATSC Technology Group 3 (TG3), known as S44, that is responsible for developing standards for the B2X system and its associated components. In addition, several ad hoc groups operate under S44 to focus on specific domains of B2X technology, including S44‑1 for Joint Architecture and Broadcast Core Network (BCN), S44‑2 for B2X RAN (System discovery and signaling, B2X Physical layer (new) and B2X Link layer protocol), and S44‑3 for B2X Systems (Interworking framework, L3 and above signaling).} The B2X System Discovery and Signaling framework, including the bootstrap signal, is currently defined in the A/392-21 specification~\cite{A39221}\footnote{The designs developed in this work are contributed to the standards by the authors of this manuscript.}. However, detailed receiver-oriented analyses of B2X bootstrap signaling and synchronization behavior remain limited in the open literature. This paper addresses this gap by examining bootstrap detection, synchronization, and receiver performance across a wide range of channel conditions and deployment environments.
\subsection{Our Contributions} 
The key contributions of this paper are summarized as follows: 
\begin{enumerate}
{
\item The design and generation of scalable Virtual Frame Start (VFS) and Slice Start (SS) bootstrap configurations to support receivers with different bandwidth capabilities, including narrowband receivers.
\item Cross-correlation analysis of the B2X and ATSC 3.0 bootstrap signals, demonstrating discrimination between the two systems by their respective bootstrap receivers.}
\item A mathematical model of the B2X receiver is developed, capturing the key signal‑processing blocks, timing relationships, and parameter dependencies required for detection and signaling decoding.
\item The estimation and correction of Integer Frequency Offset (IFO) and Fractional Frequency Offset (FFO) in the B2X receiver are analyzed, including the associated algorithms, assumptions, and performance implications across receiver processing stages.
\item A comprehensive performance analysis is conducted across a range of channel models and mobility conditions, providing detailed insight into the robustness of bootstrap-based synchronization and system discovery in B2X.
\end{enumerate}
\par These contributions provide a receiver-oriented analysis of bootstrap-based synchronization and system discovery in B2X broadcast systems.
\section{High level architecture of B2X}
\par
\begin{figure}[t!]
\centering
\vspace{-0.5cm}     \includegraphics[width=0.7\columnwidth]{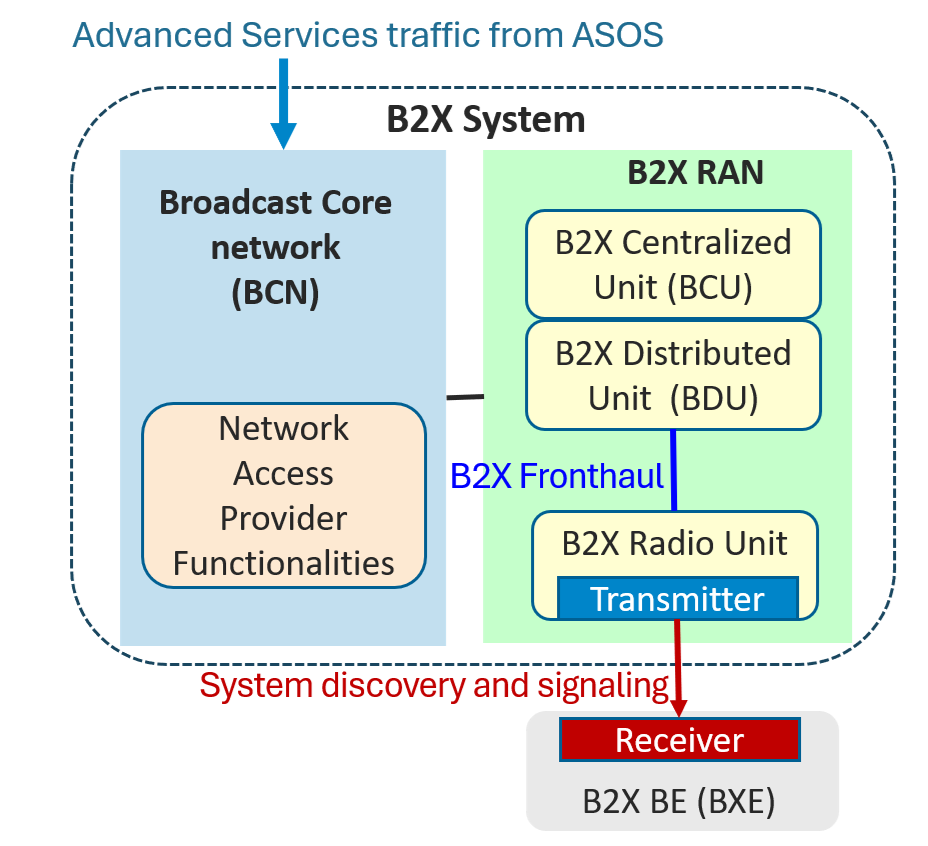}
     \vspace{-0.2cm}
      \caption{\small B2X System.}
      \label{b2x1}
      \vspace{-0.2cm}
\end{figure}

 A high-level architecture of the B2X system is shown in Fig.~\ref{b2x1}. { Please note that as the B2X system development is still in progress, this section provides the high‑level architecture context based on the currently available literature.} The B2X system comprises the Broadcast Core Network (BCN) and the B2X Radio Access Network (B2X RAN). The BCN enables Network Access Provider (NAP) functionality, among other potential deployment models. It performs key network functions, including traffic admission, billing enablement, Quality of Service (QoS) management, and bandwidth reservation. The BCN provides an interface to Advanced Service Operator Systems (ASOS), allowing an advanced service operator’s IP multicast traffic to be delivered to Broadcast Endpoints (BEs). A B2X BE (BXE) is a logical entity that receives broadcast data in the B2X system.
 \par 
\par The B2X RAN aligns with key physical-layer characteristics of the 5G RAN in order to facilitate convergence at the user plane. However, additional requirements at Layer 2 and above, as well as control-plane procedures needed to support seamless delivery of B2X and 5G traffic through the B2X RAN, remain under investigation. Through an interface defined by the BCN, traffic is delivered from the ASOS through the BCN to the B2X RAN, which processes that traffic to generate the required physical layer (PHY), Layer-1 (L1), and Layer-2 (L2) signaling prior to transmission to BEs.
\par A B2X RAN node comprises a B2X Radio Unit (BRU), responsible for lower-PHY functions and radio transmission, a B2X Distributed Unit (BDU), responsible for higher-PHY functions and B2X Baseband Processing (BBP); and a B2X Centralized Unit (BCU), which implements the B2X Link-Layer Protocol (BLP). In the present paper, however, the primary focus is the system discovery and signaling mechanism between the BRU, acting as transmitter, and the BE, acting as the receiver, using the bootstrap technique. Fig.~\ref{b2x1} illustrates the high-level architecture of the B2X system and the relationship between the core network, the B2X RAN, and broadcast endpoints.
\section{B2X Bootstrap: How it evolved from ATSC} In ATSC-based broadcast systems, a bootstrap signal is transmitted at the beginning of each frame period. It supports key receiver-side functions including initial service discovery, channel estimation, and synchronization, while also providing a mechanism for waveform identification and forward evolution. The receiver uses the bootstrap signal to determine the configuration of the waveform that follows. Within the B2X architecture, bootstrap signaling spans the BDU and BRU functional split. The bootstrap waveform is constructed at the BDU as part of baseband signal generation and delivered to the BRU through the B2X fronthaul interface using the required time-domain and frequency-domain parameters. The BRU then transmits the bootstrap signal over the radio channel to the BE. The ability of the receiver to reliably detect the information conveyed by the bootstrap signal is therefore a key design consideration for the B2X system. The following subsection provides a comparison with the ATSC~3.0 bootstrap to establish context, followed by a detailed description of the B2X bootstrap signal structure.
\subsection{Comparison with ATSC~3.0 Bootstrap Signaling}
The B2X bootstrap builds upon the system discovery and synchronization framework established in ATSC~3.0 while introducing structural and signaling enhancements to support scalable operation and alignment with an Orthogonal Frequency Division Multiple Access (OFDMA)-based physical layer. This subsection summarizes the principal similarities and differences relevant to receiver synchronization and system discovery.
\par \textit{Continuity with ATSC~3.0:}
The B2X bootstrap retains the core functional role of the ATSC~3.0 bootstrap, namely enabling system discovery, coarse time and frequency synchronization, and initial channel estimation. The underlying signal construction is preserved, based on Zadoff--Chu (ZC) sequences combined with pseudo-noise (PN) sequences and mapped onto OFDM symbols. For the normal bootstrap configuration, B2X maintains the same numerology as ATSC~3.0, including a subcarrier spacing of $\Delta\!=\!$ 3~kHz, a 2048-point Fast Fourier Transform (FFT), a 500~\!$\mu$s OFDM symbol duration, and a 4.5~MHz occupied bandwidth. The use of a prime-length ZC sequence ($N_{\mathrm{zc}} \!=\! 1499$) preserves the favorable correlation properties required for robust detection.

\par \textit{Structural evolution:}
ATSC~3.0 employs a fixed multi-symbol bootstrap located at the beginning of each frame. In contrast, B2X introduces a distributed structure consisting of VFS and SS bootstrap symbols. The VFS provides global synchronization and low-level signaling at the start of each virtual frame, while SS symbols are embedded within subframe structures to support ongoing synchronization and resource awareness. This reflects a transition from a frame-centric design to a time--frequency resource grid consistent with OFDMA operation.

\par \textit{Signaling and sequence design:}
In ATSC~3.0, bootstrap symbols are constructed in the frequency domain by the product of a PN sequence and a ZC sequence with a selected root. The product operation is designed to guarantee reflective symmetry in the frequency-domain construction. After converting to the time domain, signaling is conveyed through a relative cyclic shift between successive bootstrap symbols. In B2X, additional signaling is introduced via a controlled cyclic shift of the ZC sequence within the VFS symbols in the frequency domain. Cyclic shift values have a one-to-one mapping to Radio Frequency (RF) carrier bandwidth configurations. While this cyclic shift breaks the inherent symmetry of the ZC sequence, symmetry is restored through subcarrier mapping rules to preserve desirable detection characteristics. PN sequence handling also differs between symbol types: the PN sequence is continuous across VFS symbols, while it is reinitialized for each SS symbol, enabling flexible and distributed signaling.

\par \textit{Scalability:}
B2X extends the bootstrap framework to support operation across a range of device capabilities. In addition to the normal bootstrap ($N_{\mathrm{zc}} \!=\! 1499$), shorter ZC sequence lengths (e.g., 127, 241, 467, and 839) are defined, enabling operation with reduced sampling rates, bandwidths, and FFT sizes. This facilitates energy-efficient reception and narrowband operation while maintaining compatibility with the overall waveform structure. In contrast, ATSC 3.0 employs a fixed bootstrap configuration optimized for wideband reception.

\par \textit{Coexistence and cross-correlation considerations:}
The B2X bootstrap exhibits low cross-correlation with ATSC~3.0 signals, reducing the likelihood of false detection by receivers operating under either system. This behavior arises from differences in bootstrap structure, signaling mechanisms, and sequence construction, while preserving robust detection performance within each system. The resulting detection performance and cross-correlation characteristics are further examined in Section~\ref{VI} through B2X--ATSC cross-correlation analysis.
\par Hence, B2X preserves the robust detection framework of the ATSC~3.0 bootstrap while extending its functionality to support distributed signaling, scalable configurations, and operation within an OFDMA-based broadcast system.
\section{B2X Bootstrap signal generation}
\par This section summarizes the signal generation for the normal bootstrap configuration with $N_{\mathrm{zc}}=1499$, which serves as the baseline for receiver design and performance evaluation in this paper. Scaled bootstrap configurations are described later and follow the same underlying signal construction principles.
The B2X bootstrap signal uses a fixed sampling rate of $f_s\!=\!$ 6.144 Msamples/s and occupies a fixed bandwidth of 4.5 MHz, independent of the RF channel bandwidth used for the B2X Virtual Frame. The VFS, SS, and the L1 signaling symbols therefore occupy 4.5 MHz, whereas the user data symbols associated with a Service Category (SC) may span RF bandwidths of 5, 6, 7, 8, 10, 12, 14, 15, 16, 18, 20, 21, 24, 30, 40, or 50 MHz.
The RF carrier bandwidth is assumed to be known at the receiver. The cyclic shift of the ZC sequence in the VFS bootstrap symbols is uniquely mapped to the RF carrier bandwidth and may be used for verification. Bootstrap OFDM symbols are formed in the frequency domain using a ZC sequence multiplied by a PN sequence. 
\par
\begin{figure}[t!]
\centering
\vspace{-0.2cm}     \includegraphics[width=0.7\columnwidth]{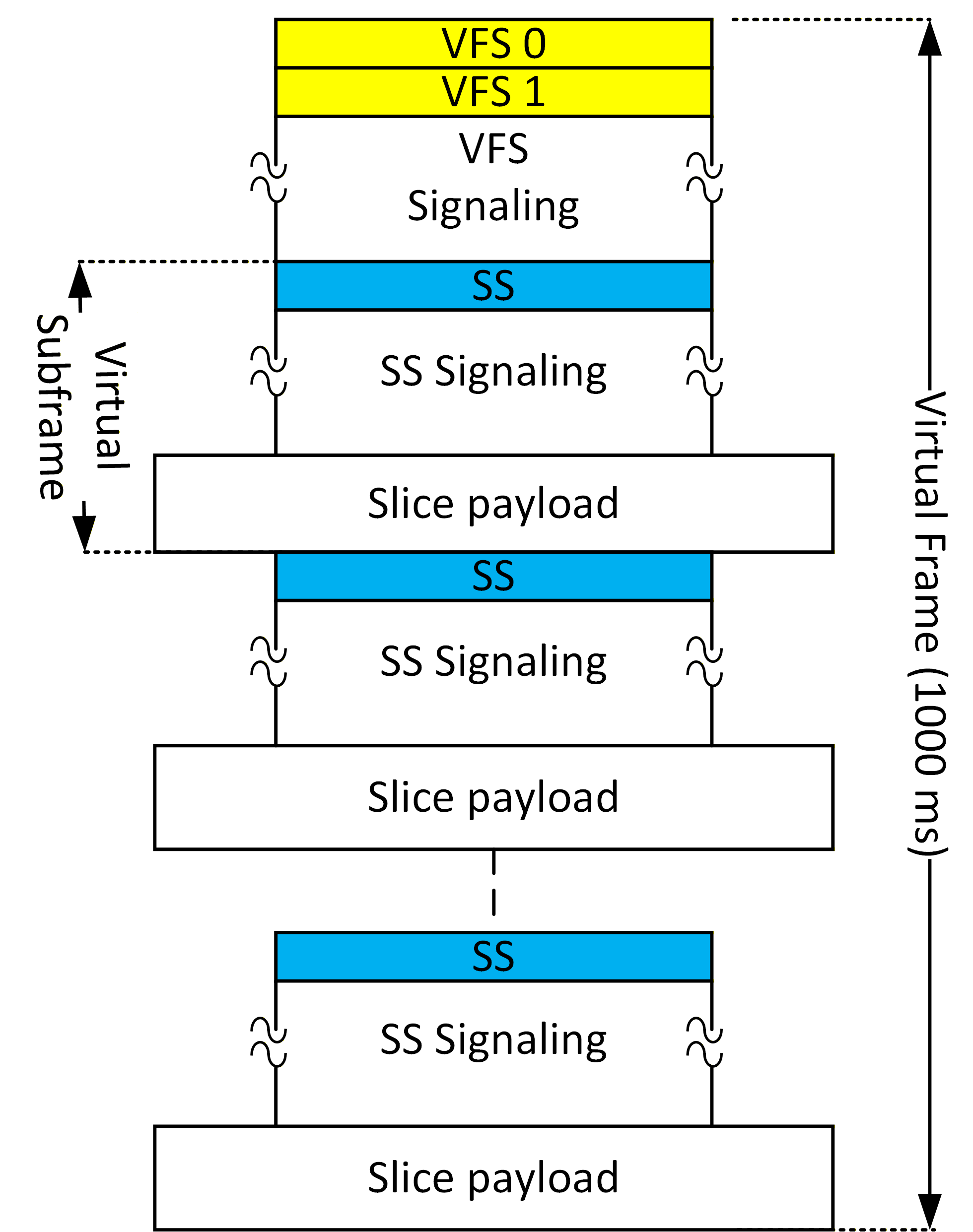}
      \caption{\small Normal B2X virtual frame.}
      \label{b2xNormalFrame}
\end{figure}

 Fig.~\ref{b2xNormalFrame} shows the normal B2X virtual frame. The B2X virtual frame duration is 1000 ms and starts with the two VFS symbols. The VFS symbols are followed by VFS Signaling symbols that carry the configuration information needed to receive the remaining virtual frame. This information includes, among other things, the number of virtual subframes, their respective duration and the service carried in each one. The VFS Signaling is followed by a variable number of virtual subframes. Each virtual subframe starts with an SS symbol, followed by SS signaling and the slice payload.

The VFS symbols are used by the B2X receiver to synchronize the virtual frame's time, frequency, and channel estimation. In addition, as described above, the second VFS symbol carries $N_b$ bits of low-level signaling. A B2X receiver interested in the service carried in a certain virtual subframe wakes up and uses the corresponding SS symbol for time and frequency resynchronization and channel estimation. The SS symbol is followed by the SS signaling, which carries information necessary to receive the slice payload. 

\subsection{VFS OFDM Symbols Generation} 
The two OFDM symbols used for the VFS bootstrap are generated in the frequency domain by multiplying a ZC sequence with a $ \pm 1$ PN sequence. The ZC sequence is cyclically shifted by an index $K$, where each value of $K$ is uniquely mapped to a corresponding RF carrier bandwidth. The ZC sequence length is $N_{\mathrm{zc}}\!=\!1499$, which is the largest prime number that supports a bandwidth within 4.5\,MHz for a subcarrier spacing of $\Delta=3$\,kHz. The ZC sequence is obtained as follows:
\begin{align}
    z_q(k) = e^{-j\pi q \frac{k(k+1)}{N_{\mathrm{zc}}}},
\end{align}
where $q\in {1,2,...,N_{\mathrm{zc}}-1}$ and $k= {0,1,2,...,N_{\mathrm{zc}}-1}$. The cyclic shift applies to the ZC sequence for two VFS (0 and 1) symbols as 
   $\tilde{z}_q(k) = z_q(k+K)$.
The maximal-length PN sequence generator is derived from a Linear Feedback Shift Register (LFSR) of length (order) $l=11$. PN sequence generation specification can be defined as $f(g,r_{\mathrm{init}})$, where $r_{\mathrm{init}}$ is the initial state (seed) of the register, and $g$ is the generator polynomial which is described as follows:
\begin{align}
   g(x) = x^{11}+x^9+1.
\end{align}
The seed of the PN sequence generator is defined in hexadecimal format. For example, if the seed is 0x136, it is converted to binary as [0001 0011 0110]. Consequently, the PN sequence starts from the least significant bit as
   $[p(0)\,\, p(1) \,\,p(2) ......\,\,p(10)] = [01101100100]$
and
    $p\left( {k + 11} \right) = p\left( k \right) \oplus p\left( {k + 2} \right)$ where $k \in \{0, 1, \ldots N_{zc} - 12\}$.
The $\pm1$ PN sequence is given by $c(k)=1-2p(k)$. The mapping of the frequency domain sequence to subcarriers is obtained by the product of the ZC and the $\pm1$ PN sequence while ensuring reflective symmetry about the DC subcarrier. Let $N_H=(N_{\mathrm{zc}}-1)/2$. For the $m$th VFS OFDM symbol, where $m \in \{0,1\}$, the $k$th subcarrier is modulated as follows:
\begin{align*}
{S_m}\left( k \right) = {\hspace{6.8cm}} \\
\left\{ {\begin{array}{*{20}{c}}
{{{\!\!\tilde z}_q}\!\left( {k + \!{N_H}} \right)c\left( {\left( {m + 1} \right){N_H}\! + k} \right)}&\!{ - {N_H}\! \le\! k \le \! - 1}\\
{{{\!\!\tilde z}_q}\!\left( {{N_H}\! - k} \right)c\left( {\left( {m + 1} \right){N_H}\! - k} \right)}&\!{1\! \le \!k \le \!{N_H}}\\
0&\!{\text{otherwise}}.
\end{array}} \right.
\end{align*}
The VFS symbols are applied to a normalized IFFT operation to yield time domain symbols as follows:
\begin{align}
\label{eq:ifftOperation}
\tilde{A}_m(n)  = \frac{1}{{\sqrt {N_{\mathrm{zc}}-1} }}\!\left(\sum\limits_{\scriptstyle k =  - {N_H}\hfill\atop
 \scriptstyle k \ne 0\hfill}^{{N_H}} {S_m}\left( k \right) e^{\left( {\frac{{j2\pi n k}}{{{N_{\mathrm{ifft}} }}}} \right)} \!\!\right),
\end{align}
where $N_{\mathrm{ifft}}  = 2048$ for the normal bootstrap, and the variance of $\tilde{A}_m(n)$ equals 1.
\begin{figure}[t!]
\centering
\vspace{-0.3cm}     \includegraphics[width=0.75\columnwidth]{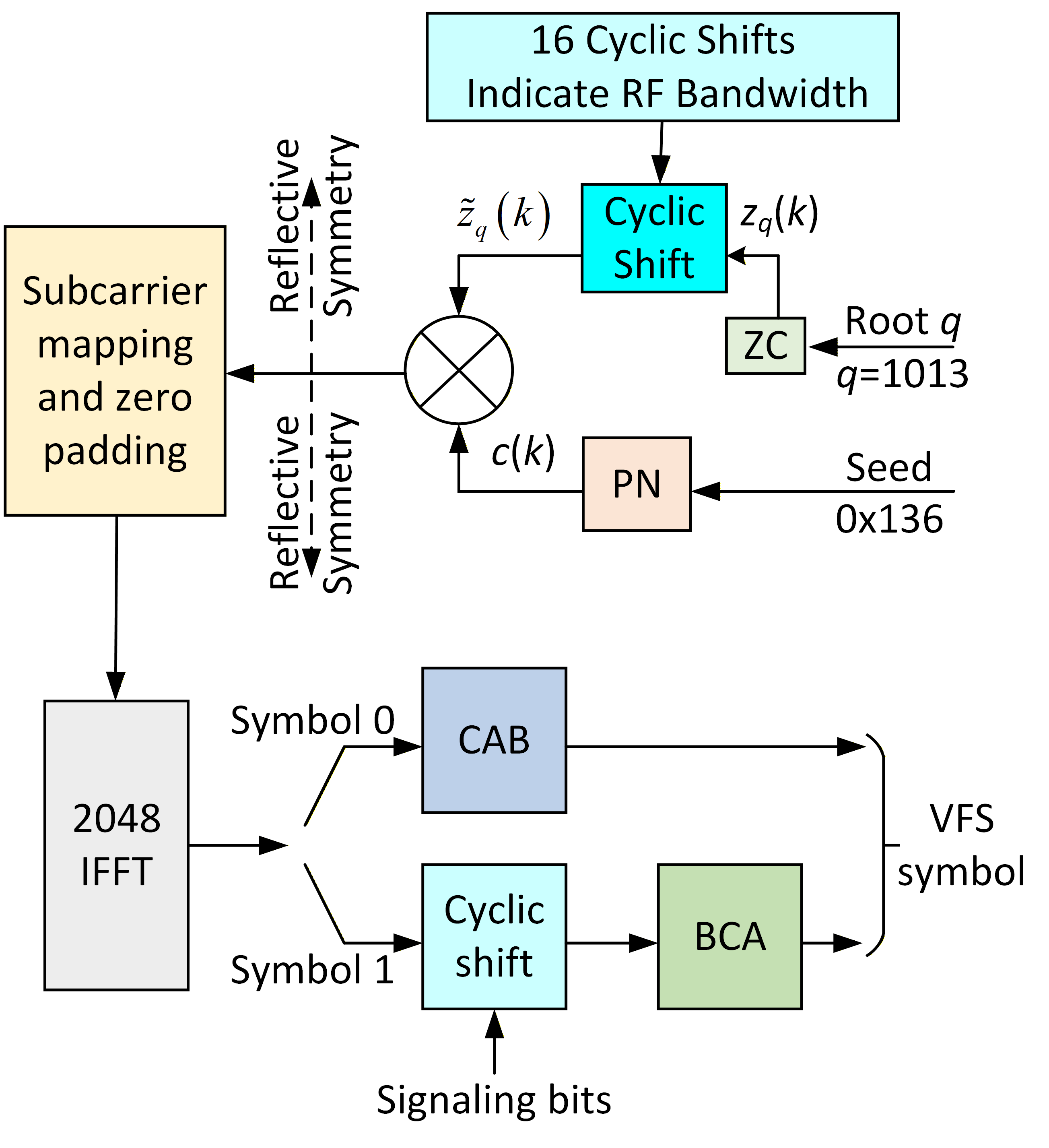}
     \vspace{-0.2cm}
      \caption{\small VFS symbols generation.}
      \label{vfsgen}
      \vspace{-0.3cm}
\end{figure}
\par Fig.~\ref{vfsgen} illustrates the generation of the two VFS symbols. The second VFS symbol undergoes an additional time-domain cyclic shift, serving as a modulation of the VFS signaling bits. Moreover, the first and second VFS symbols are extended according to CAB and BCA structures. These two operations are described next in subsections \ref{IV-c} and \ref{cab}. 

\subsection{SS OFDM Symbol Generation}
The SS symbol generation is identical to the VFS symbol 0 with two exceptions: 1) no circular shift is applied to the ZC sequence in Fig.~\ref{vfsgen}, and 2) the CAB structure parameters are different. Otherwise, the root $q$, the PN seed, the subcarrier mapping, and the IFFT operation, are the same as the VFS symbol 0. The output of the IFFT of the SS OFDM symbol is denoted as ${\tilde A_{ss}}\left( n \right)$.

\subsection{Normal bootstrap VFS signaling bits}\label{IV-c}
Signaling information is signaled via the VFS bootstrap symbol with $m=1$ through the use of cyclic shifts in the $\tilde{A}_1 (n)$ time-domain sequence. For the normal bootstrap $\tilde{A}_1 (n)$ has a length of $N_{\mathrm{ifft}} =2048$ and thus 2048 distinct cyclic shifts are possible (from 0 to 2047). With 2048 possible cyclic shifts, up to $\log_2(2048) = 11$ bits can be signaled. However, for robust signaling, only $N_b=8$ bits are used. Let $b_0,… ,b_{(N_b-1)}$ represent the values of those binary bits. Let $M (0\leq M < N_{\mathrm{ifft}} )$ represent the time domain cyclic shift for $\tilde{A}_1 (n)$. $M$ is calculated from the signaling bits using Gray code, as per the following equations. Here, $M$ is the decimal equivalent of the set of bits $m_{10} \, m_9 \, … \, m_1 \, m_0$.
\begin{equation}
\label{eq:grayCode}
    m_i=
    \begin{cases}
      \left(\sum_{k=0}^{10-i} b_k\right) \bmod\:2 , & i>10-N_b \\
      1 , & i=10-N_b \\
       0 & i<10-N_b \\
    \end{cases}
  \end{equation}
The first bootstrap symbol ($\tilde{A}_0(n)$) is not subject to time-domain cyclic shift and serves as a reference for the shift. The  cyclic shift is applied to $\tilde{A}_1 (n)$ to obtain the shifted time domain sequence from the output of the IFFT sequence as follows:
\begin{align}
\label{eq:cyclicShift}
    A_0(n) &= \tilde{A}_0(n),\nonumber \\
   A_1(n)  &= \tilde{A}_1((n+M) \bmod N_{\mathrm{ifft}} ),\\
     {A}_{ss}(n) &= \tilde{A}_{ss}(n) \nonumber.
\end{align}
Note that the cyclic shift after the IFFT operation in \eqref{eq:cyclicShift} can equivalently be implemented using a phase shift before the IFFT operation using the well-known FFT property:
\begin{align}
\label{eq:fftProperty}
    x\left( {n - M} \right) \leftrightarrow X\left( k \right)\exp \left( { - {{j2\pi Mk} \mathord{\left/
 {\vphantom {{j2\pi Mk} N}} \right.
 \kern-\nulldelimiterspace} N}} \right) \;.
\end{align}

\subsection{CAB and BCA time domain structures}
\label{cab}
Similar to \cite{standard2016system}, to support time and frequency synchronization, the bootstrap OFDM symbols are designed with special structures known as CAB and BCA. Each bootstrap symbol has three parts: A, B, and C. Part A is provided by $A_0(n)$ , $A_1(n)$, or $A_{ss}(n)$, as described in \eqref{eq:cyclicShift}, and consists of $N_A=N_{\mathrm{ifft}} $ samples. Parts B and C are composed of samples taken from part A and placed before or after part A (see Fig.~\ref{cab_bca}). This sample duplication is used for time and frequency synchronization, as described later in the paper. The total number of samples is $N_t=N_A+N_B+N_C=3N_A/2$, for an equivalent duration of 500 µs. While \cite{standard2016system} employs up to 4 {bootstrap symbols}, the B2X system employs only 2 VFS symbols.
\begin{figure}[t!]
\centering
\vspace{-0.4cm}     \includegraphics[width=0.9\columnwidth]{CAB_BCA3_updated.png}
     \vspace{-0.2cm}
      \caption{\small CAB and BCA structures.}
      \label{cab_bca}
      \vspace{-0.5cm}
\end{figure}
\vspace{-3mm}
\par Fig.~\ref{cab_bca} illustrates the two variants of the time domain symbol structure: CAB and BCA. For the two VFS symbols $N_B = N_C = N_A/4$, while for the SS symbol $N_B=31N_A/128$ and $N_C = 33N_A/128$. These values of $N_B$ and $N_C$ are sufficiently different from those used in \cite{standard2016system} to nullify the cross-correlation between the B2X VFS and the ATSC-3 {bootstrap} detection. To reduce this cross-correlation further, the frequency offset employed in the {bootstrap} of \cite{standard2016system} is not used. The value of $N_C$ is sufficiently different between the B2X VFS and SS to reduce their cross-correlation and remove any potential synchronization ambiguity between them.



\subsection{Scaled VFS and SS virtual frame} 
\begin{figure}[t!]
\centering
\vspace{-0.4cm}     \includegraphics[width=\columnwidth]{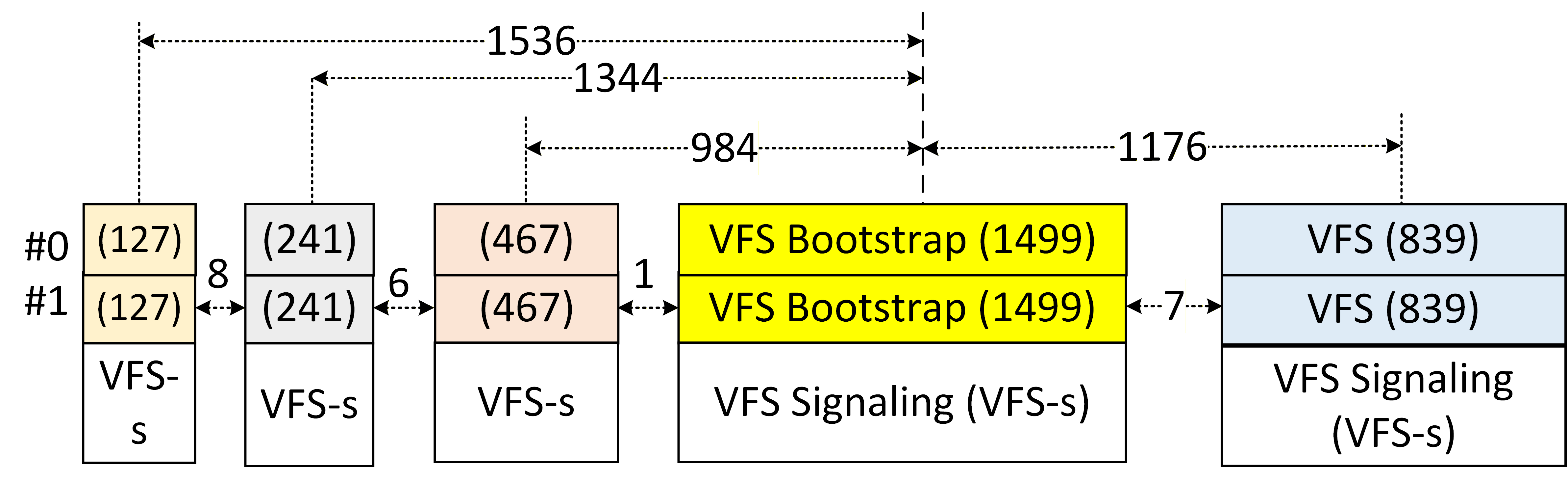}
     \vspace{-0.5cm}
      \caption{\small Scaled B2X VFS.}
      \label{b2xScaledVFS}
      \vspace{-0.3cm}
\end{figure}

The normal VFS and SS bootstraps span 1499 subcarriers. Since the subcarrier spacing is $\Delta=3$ kHz, the VFS and SS symbols' bandwidth is about 4.5 MHz. On the other hand, one of the goals of the B2X system is to support low-cost, narrowband receivers used in Internet of Things (IoT) and similar systems. Accordingly, the B2X system supports five options for scaled systems with 1, 2, 3, 4, or 5 VFS bootstraps. Fig.~\ref{b2xScaledVFS} shows the all five options, which support 5 VFS scaled bootstraps. The largest VFS bootstrap is the normal bootstrap with $N_{\mathrm{zc}}=1499$ and is placed at the center frequency. 
\par The remaining 4 VFS bootstraps are unique to the scaled B2X system. The scaled VFS bootstrap symbols are generated the same way as the normal bootstrap, with the following constraints: The ZC sequence lengths are 839, 467, 241, and 127. The corresponding ZC roots are 3, 3, 163, and 109, respectively. The PN seed is always 0x100. These values are found by simulation to provide satisfactory synchronization performance. The cyclic shift of the ZC sequence in Fig.~\ref{vfsgen}, which indicates the RF bandwidth, is not used. The IFFT size in Fig.~\ref{vfsgen} is $N_{\mathrm{ifft}} =4096$ (or higher), which allows the mapping of all 5 VFS bootstraps subcarriers. Since the subcarrier spacing is fixed at 3 kHz, the sampling frequency is increased to $f_s=12.288$ MHz. Also, the values of $N_A$, $N_B$, and $N_C$ of the CAB and BCA structures are doubled (or scaled) accordingly. The center frequencies of the 4 bootstraps are shifted from the center frequency of the normal bootstrap by 1176, -984, -1344, and -1536 subcarriers, respectively. This is shown in Fig.~\ref{b2xScaledVFS}. The number of bits carried by the second VFS symbols are $N_b=$ 7, 6, 5, and 4, respectively. Hence, the maximum number of cyclic shifts applied to the second VFS symbol is 128, 64, 32, and 16, respectively. These cyclic shifts are calculated in a similar manner to \eqref{eq:grayCode}. \par The second VFS bootstrap symbol of each scaled VFS bootstrap (including the normal bootstrap) has its own time-domain cyclic shift in Fig.~\ref{vfsgen}. To avoid using a separate IFFT operation for each scaled bootstrap, this cyclic shift is implemented in the frequency domain, i.e., before the IFFT operation. The FFT property in \eqref{eq:fftProperty} is used, where the subcarrier index $k$ includes the index of the center frequency of the respective bootstrap. The normalization factor in \eqref{eq:ifftOperation} is the total number of active subcarriers in the VFS bootstrap. For the option shown in Fig.~\ref{b2xScaledVFS}, this is the sum of all 5 ZC lengths, minus 5.
\par Similar to the normal bootstrap, SS bootstrap symbols are generated for each scaled VFS bootstrap, with the same ZC length. They serve as the first symbol in the virtual subframe. However, there are a few differences in the scaled SS bootstraps, as follows: Contrary to the normal bootstrap, the center frequency of the SS bootstrap may not be the same as the center frequency of the associated VFS bootstrap. The VFS Signaling may indicate that the associated SS bootstrap symbol is at a different center frequency. This provides flexibility and higher efficiency in the spectrum utilization. However, the B2X receiver has to retune its frequency after decoding the VFS Signaling. Moreover, a VFS bootstrap may be associated with multiple SS bootstrap symbols with the same bandwidth at different center frequencies, i.e., at different parts of the band. This provides the flexibility to deliver more services across different parts of the band with the same narrow bandwidth. 

Since the main structure of the normal and scaled VFS and SS bootstrap symbols is the same, the same baseband receiver algorithm designed for the normal bootstrap can be readily used for any of the scaled bootstraps. The scaled bootstrap is first down-converted to zero frequency, filtered according to its bandwidth, and down-sampled. After down-sampling\footnote{The down-sampled rate satisfies the Nyquist criterion; therefore, synchronization and decoding performance are not compromised. Furthermore, it reduces the number of FFT computations and the overall digital processing requirements.}, the same base-band receiver algorithms can be used for any of the 5 bootstrap bandwidths in Fig.~\ref{b2xScaledVFS}.    

\begin{figure*}[t!]
\centering
\vspace{-0.4cm}
     \includegraphics[width=1.8\columnwidth]{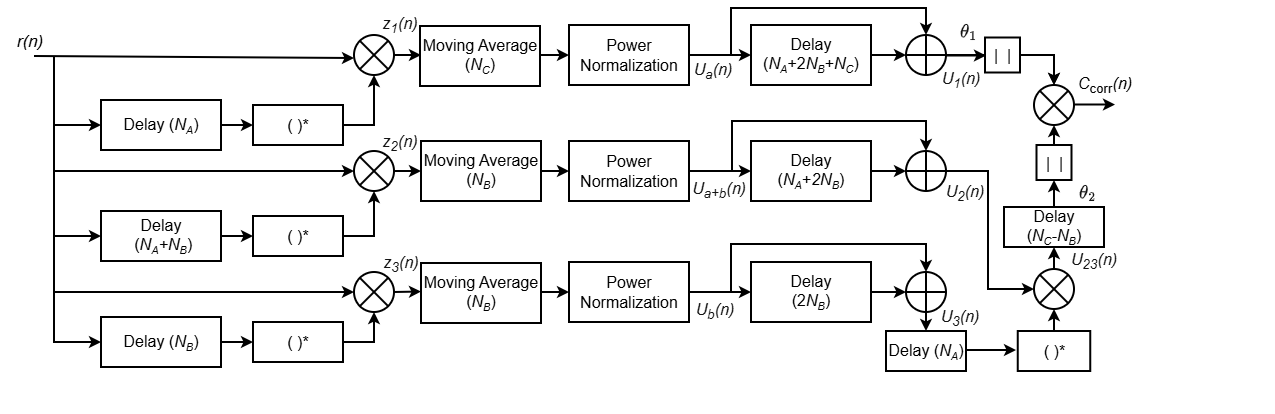}
     \vspace{-0.3cm}
      \caption{\small Delayed correlation-based detection and timing estimation.}
      \vspace{-0.4cm}
      \label{delayedCorrelation}
\end{figure*}

\section{B2X Receiver Design}
{We develop a delayed-correlation receiver for detecting the B2X bootstrap signal, motivated by its simple structure and suitability for robust synchronization under different channel and mobility conditions. As discussed in~\cite{He_2016}, bootstrap detection can also be performed using a Local Sequence-Based Correlation (LSBC) receiver; however, its suitability may depend on the propagation environment, and its implementation involves multiple differential operations, resulting in slightly higher computational complexity. Therefore, we adopt the delayed-correlation approach in this work.
The B2X receiver is based on the ATSC 3.0 bootstrap receiver in~\cite{He_2016}, with modifications to accommodate the B2X bootstrap structure. Specifically, the receiver in~\cite{He_2016} uses four bootstrap symbols, whereas B2X uses only two for synchronization, thereby requiring a modified delayed-correlation structure. In addition, the initial phase rotation operation used in~\cite{He_2016} is not required in the B2X receiver.} Detection starts with timing estimation of the first bootstrap symbol by processing delayed correlations, followed by FFO estimation using phase value(s) of delayed correlation signals. This is followed by IFO estimation using frequency-domain correlation over a search window of all possible IFOs, selecting the best match in a maximum-likelihood (ML) fashion. These estimated FFO and IFO values are used to compensate for the frequency offset. After the frequency correction, detection validation is performed by calculating errors in decoding signaling bits. Further parameter decoding is performed to recover intended signaling parameters.
\par
\begin{figure}[t!]
\centering
\vspace{-0.2cm}     \includegraphics[width=0.9\columnwidth]{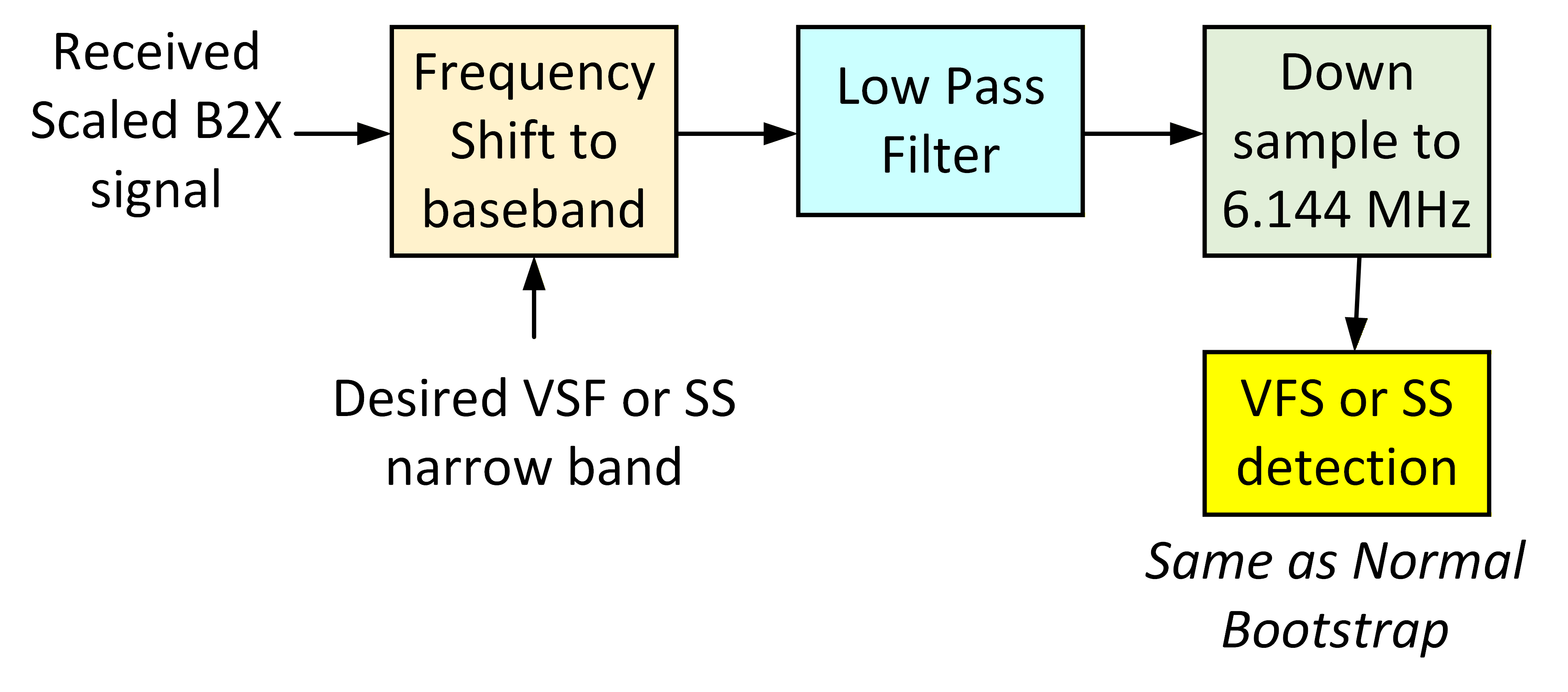}
     \vspace{-0.4cm}
      \caption{\small Scaled bootstrap detection via normal bootstrap algorithm.}
\label{scaledBootstrapDetection}
      \vspace{-0.4cm}
\end{figure}

 In the following, we focus on the receiver design for the normal bootstrap, i.e., $N_{\mathrm{zc}}=1499$ and $f_s=6.144$ MHz. To detect any of the other 4 narrowband VFS bootstraps in Fig.~\ref{b2xScaledVFS}, the desired narrowband bootstrap is first converted to zero frequency, low-pass filtered, downsampled, and applied for detection processing. To employ the same baseband processing for any narrowband VFS, it is down-sampled from 12.288 MHz to 6.144 MHz and processed the same way as the normal bootstrap, as described below. This is shown in Fig.~\ref{scaledBootstrapDetection}. 

\subsection{Delayed correlation-based detection and Timing estimation} \label{IVa}
{ The} delayed correlation-based receiver design is shown in Fig.~\ref{delayedCorrelation}. It performs detection and timing estimation for the two-symbol VFS bootstrap employed in B2X. As shown in Fig.~\ref{delayedCorrelation}, the receiver uses the CAB and BCA structures. It consists of three delay branches with delays $N_A$, $N_A+N_B$, and $N_B$, respectively. Correlation signals are denoted as $U_a(n)$, $U_{a+b}(n)$, and $U_b(n)$, which are further accumulated with delays by $N_A+2N_B+N_C$, $N_A+2N_B$, and $2N_B$, respectively, to combine the correlation peaks due to the two VFS bootstrap symbols. The combined correlations are denoted as $U_1(n)$, $U_2(n)$, and $U_3(n)$. The conjugate of $U_3(n)$ is multiplied by $U_2(n)$ to enhance the peak value, and results in $U_{23}(n)$. The final correlation signal is obtained as  follows:
\begin{align}
   C_{\mathrm{corr}}(n) = |U_{23}(n+(N_C-N_B))|*|U_1(n)|.
\end{align}
Note that in B2X $N_C\!\!=\!\!N_B$, which considerably simplifies the system compared to ATSC~3. The peaks of $U_{23}(n)$ and $U_1(n)$ are aligned, so the output peak of $C_{\mathrm{corr}}(n)$ is strong enough to be located accurately. The timing estimation is obtained by locating the peak of the $C_{\mathrm{corr}}(n)$ as $\hat{n} = \arg \max_n \{{ C_{\mathrm{corr}}(n)}\}$.
The starting sample of first VFS frame $0$ is estimated at time instant $n_{0}$ as $n_{0}= \hat{n} - 2 N_t$,
where $\hat{n}$ is the time instant at the peak of $C_{\mathrm{corr}}(n)$ and $N_t= N_A+N_B+N_C$.

\subsubsection{Mathematical formulation of the receiver} 
{At the receiver input, the impaired time-domain discrete-time signal $r(n)$ can be expressed as
\begin{align}
\label{eq:receivedSignal_actual}
r(n) = e^{\frac{j2\pi n f_o}{f_s}}
\left( s_t(n) \ast h(n) \right)
+ w(n),
\end{align}
where $s_t(n)$ denotes the transmitted signal, $f_o$ is the frequency offset, $h(n)$ is the faded channel, and $w(n)\sim\mathcal{CN}(0,1)$ denotes normalized circularly symmetric complex additive white Gaussian noise (AWGN).}
\par To clarify the receiver processing, we consider a noiseless and fadeless version of \eqref{eq:receivedSignal_actual}, denoted by $r_0(n)$, as follows:
\begin{align}
\label{eq:receivedSignal}
   r_0(n) = e^{\frac{j2\pi n f_o }{f_s}} s_t(n).
\end{align}
 Referring to the top branch in Fig.~\ref{delayedCorrelation}, the multiplication of the received signal with the complex-conjugate, delayed version of itself yields $z_1(n)$ as follows:
\begin{align*}
z_1(n)&=  e^{\frac{j2\pi N_A f_o }{f_s}} s_t(n) s_t^*(n-N_A).
\end{align*}
After passing $z_1(n)$ through the moving average filter and power normalization, $U_a(n)$ can be approximated as two successive impulses\footnote{The practical correlation output may contain additional weak correlation components, as observed in Fig.~\ref{add3}; however, only the dominant correlation peaks are retained in the following analysis for mathematical tractability.} (assuming the starting point is $n=0$):
\begin{align*}
U_a(n)&= e^{\frac{j2\pi N_A f_o }{f_s}} (\delta(n-N_A-N_C) \\&+ \delta(n-2N_A-2N_B-2N_C))
\end{align*}
The two impulses $U_a(n)$ are due to the cross-correlation of the $N_C$ samples in the first and second VFS bootstrap symbols, respectively.
Accordingly, $U_1(n)$ coherently combines the two strong impulses in $U_a(n)$, to yield $U_1(n)$ as follows:
\begin{align}
U_1(n) &= U_a(n)+U_a(n-N_A-2N_B-N_C),\nonumber\\
&= e^{\frac{j2\pi N_A f_o }{f_s}} (\delta(n-N_A-N_C) \nonumber\\&+ 2\delta(n-2N_A-2N_B-2N_C) \nonumber\\&+ \delta(n-3N_A-4N_B-3N_C) ).
\end{align}
The mathematical expression for $U_1(n)$ is also verified by observing the peaks in the simulation results, as shown in Fig.~\ref{add3}, Section~\ref{VII-C}.\\
In the middle correlation branch of Fig.~\ref{delayedCorrelation}, the multiplication of $r_0(n)$ with the delayed complex-conjugate of itself yields $z_2(n)$ as:
\begin{align*}
z_2(n)&=  e^{\frac{j2\pi (N_A+N_B) f_o }{f_s}} s_t(n) s_t^*(n-N_A-N_B).
\end{align*}
After applying the moving average filter and power normalization on $z_2(n)$, $U_{a+b}(n)$ can be approximated as a combination of two impulses. Each impulse is due to the match of $N_B$ samples within the first and second VFS bootstrap symbols, respectively. Further, $U_{a+b}(n)$ can be obtained as:
\begin{align*}
U_{a+b}(n)&= e^{\frac{j2\pi (N_A+N_B) f_o }{f_s}} (\delta(n-N_A-N_B-N_C) \\&+ \delta(n-2N_A-3N_B-N_C)).
\end{align*}
Accordingly, $U_2(n)$ can be expressed as follows:
\begin{align}
U_2(n) &= U_{a+b}(n)+U_{a+b}(n-N_A-2N_B) \nonumber\\
       &= e^{\frac{j2\pi (N_A+N_B)f_o}{f_s}}
       \Big(
       \delta(n-N_A-N_B-N_C) \nonumber\\
       &\quad +2\delta(n-2N_A-3N_B-N_C) \nonumber\\
       &\quad +\delta(n-3N_A-5N_B-N_C)
       \Big). \label{eq:U2_expanded}
\end{align}
Similarly, in the third delayed correlation branch with a delay of $N_B$, the output of the multiplier results in $z_3(n)$ as follows:
\begin{align*}
z_3(n)&=  e^{\frac{j2\pi N_B f_o }{f_s}} s_t(n) s_t^*(n-N_B).
\end{align*}
After passing through the moving average filter and power normalization, $U_{b}(n)$ can be approximated as a combination of two impulses:
\begin{align*}
U_{b}(n)&= e^{\frac{j2\pi N_B f_o }{f_s}} (\delta(n-N_A-N_B-N_C) \\&+ \delta(n-N_A-3N_B-N_C)).
\end{align*}
Accordingly, $U_3(n)$ can be expressed as follows:
\begin{align}
U_3(n) &= U_{b}(n)+U_{b}(n-2N_B),\nonumber\\
&= e^{\frac{j2\pi N_B f_o }{f_s}} (\delta(n-N_A-N_B-N_C) \nonumber\\&+ 2\delta(n-N_A-3N_B-N_C) \nonumber\\&+ \delta(n-N_A-5N_B-N_C)).
\end{align}
Further $U_{23}(n)$ can be obtained as follows:
\begin{align}
U_{23}(n) &= U_2(n) \; U_3^*(n-N_A),\nonumber\\
&=4 e^{\frac{j2\pi N_A f_o }{f_s}} \delta(n-2N_A-3N_B-N_C).
\end{align}
Hence, final expression for $C_{\mathrm{corr}}(n)$ can be obtained as (note that $N_B=N_C=N_A/4$ for the VFS):
\begin{align}
   C_{\mathrm{corr}}(n) &= |U_{23}(n-(N_C-N_B))|\times|U_1(n)|,\nonumber\\
    &= 8e^{\frac{j2\pi (2N_A) f_o }{f_s}} \delta(n-2N_t).
\end{align}
The final output $C_{\mathrm{corr}}(n)$ shows one main peak at $2N_t$, which is further boosted by delayed correlations in the receiver. All other peaks are diminished after the multiplications.
All peaks are also verified using MATLAB simulations, as shown in Fig.~\ref{add3} and Fig.~\ref{finalCorr}, presented in Section \ref{VII-C}.

\section{Synchronization and decoding of bootstrap signal at receiver}
\subsection{FFO and IFO estimation/correction} 
The carrier frequency offset (CFO) is treated as the sum of an FFO component and an IFO component.
\par FFO can be estimated from the angles $\theta_1$ and $\theta_2$ in Fig.~\ref{delayedCorrelation}. These angles are captured at the final peak location estimated as explained above. Using $U_1(n)$ and $U_{23}(n)$ an angle $\theta_{\mathrm{ref}}$ is calculated, and the FFO (fraction of the subcarrier spacing $\Delta$), denoted as $\epsilon$, is derived as follows:
\begin{align}
\label{eq:FFO}
  \theta_{\mathrm{ref}} &= \operatorname{angle}\left(  U_1(2N_t) + U_{23}(2N_t)   \right) \\
  \label{eq:FFO2}
  \epsilon &=\frac{\theta_{\mathrm{ref}}f_s}{2\pi N_A \Delta}
 \end{align}
As it can be observed that $U_1(2N_t)$ and $U_{23}(2N_t)$ add coherently to provide a better estimation of the FFO. 

\par Once the timing estimation and the FFO estimation are completed using the delayed correlation detector, the two VFS symbols and the SS symbol(s) are captured from the received signal in \eqref{eq:receivedSignal_actual}. The time-domain location of the SS symbols is provided by the signaling message following the two VFS bootstrap symbols, as shown in Fig.~\ref{b2xNormalFrame}. The processing of the SS symbols is described later in the paper. The $2N_t$ samples of the VFS symbols are FFO corrected using the estimated FFO in \eqref{eq:FFO2}, to use for IFO estimation. The VFS samples are multiplied by $e^{\frac{-n\theta_{\mathrm{ref}}}{N_A}}$, $0 \le n < 2N_t$. Subsequently, and referring to the CAB and BCA structures in Fig.~\ref{cab_bca}, the $N_A$ samples of the A parts of the first and second received VFS symbols, respectively, are extracted for IFO estimation. Below, we assume that the IFO is within $\pm L \Delta$ Hz, where $L$ is an integer.

\par FFT is applied to both symbols to yield the two received, FFO corrected, frequency domain VFS symbols $Z_0(k)$ and $Z_1(k)$, $0 \le k < N_A$. The two symbols $Z_0(k)$ and $Z_1(k)$ are used together with the transmitted VFS symbols $S_0(k)$ and $S_1(k)$ which are inputs of \eqref{eq:ifftOperation} to estimate the IFO $l$, $-L \le l \le L$. A metric $V_l$ is calculated for each possible IFO value $l$, and the largest value of $V_l$ is selected.

\par Denote $S_m(k-l)$ as the cyclic shift of the transmitted (reference) symbol $S_m(k)$, $m=0, \; 1$. The IFO is estimated as follows: 
\begin{enumerate}
\item The received symbol $Z_m(k)$ is multiplied by the cyclic shifted, reference symbol $S_m(k-l)$ to yield $D_m(k,l)$ as:
\begin{equation}
    D_m(k, l) = Z_m(k) S_m^*(k-l) \: \: \: , \: \: m=0,1. \nonumber
\end{equation}
\item The frequency domain vector $D_m(k, l)$ is applied to IFFT of size $N_A$ and the maximum absolute value of the IFFT output $Y_m(l)$ is found. 
\item Using the two VFS symbols, the metric $V_l$ is given by $V_l = |Y_0(l)|^2 + |Y_1(l)|^2$ 
\item Steps (1), (2) and (3) are repeated for each possible IFO value $l$. The estimated IFO is given by the value of $l$ corresponding to the maximum value of $V_l$, $-L \le l \le L$.
\end{enumerate}
\par The total CFO in Hz is given by $\hat{f}_0(l+\epsilon) \Delta$ , where $\epsilon$ is given by \eqref{eq:FFO2}. The two VFS symbols (i.e, $2N_t$ samples) and the SS symbol ($N_t$ samples) are re-captured from the received signal in \eqref{eq:receivedSignal_actual} and their full CFO is corrected. Subsequently, after CFO correction, the $N_A$ samples of the A parts of the first and second received VFS symbols, respectively, are re-extracted for channel estimation and cyclic shift detection, as described below. The CFO-corrected VFS symbols are applied to FFT to yield $\hat{Z}_m(k)$, $m=0,1$.

\subsection{Channel estimation and cyclic shift decoding} \label{VI-B_CEandDecoding}
The time-domain cyclic shift is computed by performing channel estimation using locally generated reference signals. First, both channel estimates are found using locally generated reference signals as follows:
\begin{align}
H_0(k) &= \hat{Z}_0(k)S_0^*(k), \\
H_1(k) &= \hat{Z}_1(k)S_1^*(k), k = 0, 1, \ldots, N_A - 1
\end{align}
where $\hat{Z}_0(k),\hat{Z}_1(k)$ are the CFO corrected VFS symbols in the frequency domain and $S_0(k), S_1(k)$ are the locally generated reference VFS symbols in the frequency domain, which are inputs of~\eqref{eq:ifftOperation}. {Thus, the channel estimation is equivalent to zero-forcing, with $S_0(k)$ and $S_1(k)$ serving as reference signals.}

After estimation, signaling bits decoding is performed by finding the maximum index of the combined (multiplied) channel estimations (after converting to the time domain) as follows:
\begin{align}
H(k)&=H_0(k)H_1^*(k),\nonumber\\
\hat{M} &= \arg [\max{\{|\text{ifft}{(H(k))|\}}}].
\end{align}
Further $\hat{M}$ can be represented as binary values $[m_{10} m_9 ...m_1 m_0]$ from which signaling bits can be extracted as follows:
\begin{equation}
\label{eq:signaling_bit_dec}
    b_j=
    \begin{cases}
    m_{10} & j=0\\
     m_{11-j} \oplus m_{10-j} & 1\leq j \leq N_b \\
       0 & N_b\leq j < 11 \\
    \end{cases}
  \end{equation}
{If one or more bits of the frame are in error, the entire frame is considered erroneous. By counting the number of frames in error, the probability of frame error ($P_e$) can be defined as the ratio of frames in error to the total number of transmitted frames.}
\subsection{SS detection} 
As illustrated in Fig.~\ref{b2xNormalFrame}, each payload slice in the virtual frame is preceded by an SS symbol. As explained above, after time synchronization, the SS symbol(s) location is provided by the signaling message after the two VFS symbols. However, since the frame may be long, and due to channel fading and sampling frequency, this location may be slightly offset. Hence, the SS symbols provide a means of resynchronization for the associated payload slice. The $N_t$ samples of each SS bootstrap symbol are first CFO corrected. Then, the $N_A$ samples of part A of the SS bootstrap symbol are captured (it is recommended to back off by $B$ samples and capture the $N_A$ samples starting $B$ samples in the C part of the CAB structure). Using circular cross-correlation, the true start of part A of the SS bootstrap is found.

\par Define the CFO corrected received SS bootstrap samples as $r_{ss}(n)$, $n=0, 1, ..., N_A-1$. To search for resynchronization over a window $-W \le m \le W $, a normalized circular cross-correlation $\widetilde{C}_{\mathrm{corr}}(m)$ is performed as:
\begin{equation}
\label{eq:sliding_corr}
\widetilde{C}_{\mathrm{corr}}(m) = \frac{{\sum\limits_{n = 0}^{{N_A} - 1} {{r_{ss}}\left( {n - m} \right)A_{ss}^*\left( n \right)} }}{{\sqrt {{N_A}\;\sum\limits_{n = 0}^{{N_A} - 1} {{{\left| {{r_{ss}}\left( n \right)} \right|}^2}} } }} \: \: \: ,
\end{equation}
where $A_{ss}(n)$ are the transmitted time domain SS samples, given by \eqref{eq:cyclicShift}. The maximum absolute value of \eqref{eq:sliding_corr} provides the new synchronization for this SS, if it exceeds a predefined threshold. 

\subsection{Scalable Bootstrap detection}
One of the main features of the B2X system is its flexibility in bandwidth utilization. The scalable version of the B2X payload slices is OFDMA-based. It allows B2X devices of various bandwidth capabilities to join the system. The detailed frame format of the B2X system is a subject of a future paper. Here, we focus on the scalable bootstrap that enables system entry for both narrowband and wideband devices. Fig.~\ref{b2xScaledVFS} shows the full-fledged version of the VFS scalable bootstrap. It permits devices with up to 5 bandwidth capabilities to join the B2X system and receive a service compatible with their capabilities. Subsets of the full-fledged system are also possible.

\par To simplify the receiver's design, the narrower VFS and SS bootstraps are scaled versions of the standard bootstrap. Consequently, the same VFS and SS detection algorithms can be used at the receiver. For example, the full-fledged VFS bootstrap system requires an IFFT size of $N_{\mathrm{ifft}}  \ge 4096$. With subcarrier spacing of $\Delta=3$ kHz, the sampling frequency at the transmitter is $f_s \ge 12.288$ MHz. A receiver that desires to synchronize with any particular VFS bootstrap in Fig.~\ref{b2xScaledVFS} can simply down-convert this VFS bootstrap to zero frequency and apply it to an appropriate low-pass filter. The filter output is downsampled to 6.144 MHz and applied to the same standard VFS detection algorithm. This is shown in Fig.~\ref{scaledBootstrapDetection}. The same concept applies to the SS bootstrap symbol(s). The only difference is that the SS bootstrap symbol, with the same bandwidth as the VFS symbols, may be located at any center frequency in the frame bandwidth. As outlined earlier in the paper, the total RF bandwidth of the B2X frame can range from 5 MHz to 50 MHz. The SS bootstrap and the subsequent payload can be placed at any center frequency, as specified by the VFS signaling message.

\par It is important to note that while the filter in Fig.~\ref{scaledBootstrapDetection} should satisfy the Nyquist criterion, it does not need to have a tight bandwidth around the desired VFS bootstrap. If the filter allows some leakage from the neighbor bandwidth VFS bootstrap(s), this leakage can actually enhance VFS synchronization, since the CAB and BCA structures of all VFS-scaled bands are identical. Hence, the delayed correlator in Fig.~\ref{delayedCorrelation} is not affected. A similar argument applies to the SS bootstrap(s).

\section{Design Associated Findings} \label{VI}
This section highlights key observations associated with the B2X bootstrap and its role in system discovery and signal differentiation, with emphasis on signal structure, detection behavior, and system-level interactions.

\par \textit{ZC--PN sequence design:}
The selection of ZC root and PN sequence parameters was evaluated over a large design space, with only minor differences observed in detection and decoding performance. The B2X bootstrap uses full-length PN sequences, which preserve the correlation properties of the underlying m-sequence, whereas truncation, as used in ATSC~3.0, degrades these properties. The impact of this difference is not isolated in the results presented here, but the correlation structure of the bootstrap signal is maintained.

\par \textit{Detection and false alarm behavior:}
Bootstrap detection is primarily driven by time-domain autocorrelation of repeated signal components. As a result, detection behavior is strongly influenced by the CAB/BCA structure rather than the specific Zadoff--Chu or PN sequence parameters, whose impact on detection performance is comparatively limited. Evaluations indicate that changes to the time-domain structure introduce a trade-off between reliable B2X detection and reduced likelihood of detection by legacy ATSC~3.0 receivers. The resulting B2X bootstrap structure reflects this trade-off and supports distinguishability across different bootstrap signal configurations.

\par \textit{Frequency shift and interference robustness:}
Prior systems' bootstrap construction includes a frequency shift to mitigate the possible presence of continuous-wave interference (CWI). The impact of this frequency shifting within the B2X bootstrap structure was examined. Analysis and simulation results indicate limited benefit for the frequency shift under practical receiver configurations. Removing the frequency shift does not degrade detection performance, including false-alarm behavior, while simplifying the bootstrap structure.

\par \textit{Scalable bootstrap characteristics:}
In scalable bootstrap configurations, multiple bandwidth parts are closely spaced in frequency, and receiver filtering may include contributions from adjacent bands. Since all bandwidth parts share a common time-domain structure, these contributions remain aligned and can support the autocorrelation process used for synchronization. Detection performance remains largely insensitive to specific ZC root and PN sequence choices, while selection of PN sequences primarily influences false alarm behavior between service categories. This reflects the importance of consistent time-domain structure across bandwidth parts in enabling scalable operation.

\par \textit{B2X--ATSC cross-correlation characteristics:}
Operation across both broadcast and IMT spectrum bands may introduce scenarios in which receivers encounter signals from different systems. Cross-correlation behavior between ATSC~3.0 and B2X bootstrap signals is therefore examined to assess the robustness of system discovery under such conditions. Specifically, the response of a B2X receiver to an ATSC~3.0 bootstrap signal, and the response of an ATSC~3.0 receiver to a B2X bootstrap signal, are evaluated. 

\par
\begin{figure}[t!]
\centering
\vspace{-0.2cm}
\includegraphics[width=\columnwidth]{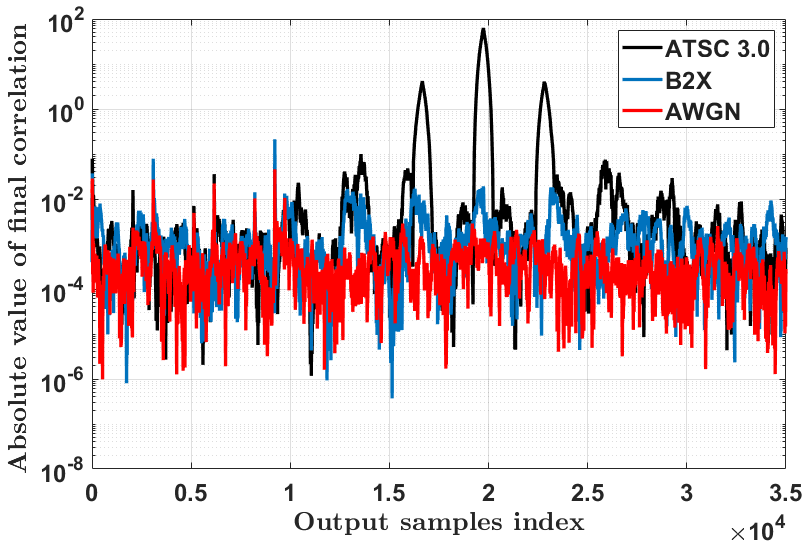}
\vspace{-0.5cm}
\caption{\small ATSC 3.0 delayed correlator output, when inputs applied (i) ATSC 3.0 (ii) B2X and (iii) AWGN only.}
\label{ATSCinputPoint3}
\vspace{-0.3cm}
\end{figure}

\begin{figure}[t!]
\centering
\vspace{-0.2cm}
\includegraphics[width=\columnwidth]{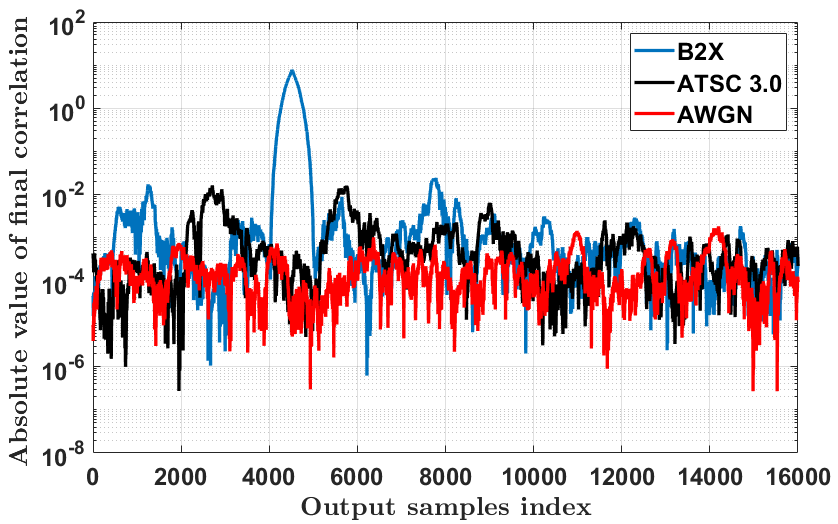}
\vspace{-0.5cm}
\caption{\small B2X delayed correlator output, when inputs applied (i) B2X (ii) ATSC 3.0 and (iii) AWGN only.}
\label{B2XandATSC_InB2X}
\vspace{-0.3cm}
\end{figure}

 When applying noiseless versions of both the ATSC~3.0 bootstrap and the B2X bootstrap signal to the ATSC~3.0 delayed correlator receiver in \cite{He_2016}, the final correlation peak due to the ATSC~3.0 bootstrap is at least 438 times higher than the B2X bootstrap as shown in Fig.~\ref{ATSCinputPoint3}. Similarly, when the two bootstrap signals are applied to the B2X delayed correlator of Fig.~\ref{delayedCorrelation}, the final correlation peak due to the B2X bootstrap is at least 267 times higher than the ATSC~3.0 bootstrap, as shown in Fig.~\ref{B2XandATSC_InB2X}. This confirms the ability of each delayed correlator to reject the undesired bootstrap. The ATSC~3.0 delayed correlation is more capable of suppressing the B2X bootstrap, since it combines correlations across  { 4 bootstrap symbols}, whereas the B2X delayed correlator combines only 2 VFS bootstrap symbols.

\par \textit{VFS bootstrap comparison with ATSC~3.0:}
Comparative evaluations indicate that bootstrap detection and decoding performance in B2X is generally similar to that of ATSC~3.0, with modest differences observed under certain conditions. These differences are consistent with the use of fewer bootstrap symbols in B2X, which reduces time-domain averaging relative to ATSC~3.0 while enabling a more compact signaling structure.

\section{Performance analysis}
This section presents a quantitative evaluation of the B2X normal bootstrap ($N_{\mathrm{zc}}\!\!=\!\!1499$) receiver under a range of channel and mobility conditions. The results complement the design observations presented in Section \ref{VI} by characterizing detection performance, synchronization accuracy, and robustness across practical deployment scenarios.
\subsection{Channels modeling with mapping to use cases} 
A unified set of channel models as shown in Table \ref{channels}, is selected to ensure credible and comparable evaluation of B2X normal bootstrap performance across diverse broadcast deployment scenarios. The modeling spans baseline, mobile, portable, and fixed reception conditions, with mobility ranging from pedestrian to extreme high‑speed operation and with propagation environments representative of urban, rural, SFN, and rooftop installations. Standardized Tapped‑Delay‑Line (TDL) and field‑measured profiles are employed to capture realistic multipath and Doppler dynamics. Synchronization impairments such as CFO and Sampling-Frequency-Offset (SFO) are consistently applied to stress receiver robustness under practical oscillator mismatches. Overall, this channel selection provides sufficient diversity in fading, delay spread, and Doppler effects to validate bootstrap reliability under realistic B2X operating conditions.

\begin{table}[t!]
\centering
\caption{Channel models, mapped use cases, and associated parameters used for B2X simulations}
\label{channels}
\begin{tabular}{|p{0.14\textwidth}|p{0.055\textwidth}|p{0.055\textwidth}|p{0.05\textwidth}|p{0.05\textwidth}|}
\hline
\textbf{Use Case} & \textbf{Channel Model} & \textbf{\{CFO (kHz), SFO 
(ppm)\}\cite{He_2016}
} & \textbf{Speed (km/h)} & \textbf{RMS Delay Spread (ns)}\\ \hline
Performance benchmarking & AWGN & \{0,0\} & - & - \\ \hline
Typical urban (considered in legacy mobile networks)   & TU6  & \{0,0\}, \{200,30\}  & 30, 60, 120 & -
\\ \hline
Rural \& Urban (based on field measured profile for SFN environment)  & India-R and India-U & \{0,0\} & 30, 60, 120 & - \\ \hline
High Speed Train (HST) scenario & HST-SFN    & \{0,0\}, \{200,30\}  & 500 & - \\ \hline
Mobile vehicular & TDL-E    & \{0,0\}, \{200,30\}  & 30, 60, 120 &  50, 1000 \\ \hline
Indoor portable & TDL-B     & \{0,0\}, \{200,30\}  & 5 & 50, 300  \\ \hline
Fixed rooftop
Considering Non-Line-of-Sight (NLOS) & TDL-C    & \{0,0\}, \{200,30\}  & 0 & 50, 300, 800 \\ \hline
Fixed rooftop
Considering Line-of-Sight (LOS) & TDL-D    & \{0,0\}, \{200,30\}  & 0 & 50, 100, 500 \\ \hline
\end{tabular}
\vspace{-0.5cm}
\end{table}
\subsection{Simulation parameters} 

 The simulation parameters are derived from the channel configurations summarized in Table \ref{channels} and are selected to capture realistic mobility and propagation-induced impairments. TDL channels employ tap delays and powers as specified in 3GPP TR38.901, Sec.7.7.2 \cite{3gpp_tr38901_v18}. The TU-6 channel follows the COST 207 model \cite{cost207_1989}, and the India-R \& India-U channels are adopted from field-measured profiles reported in \cite{ahn2022characterization}. The Doppler spectrum is modeled using the classical Jakes formulation. All simulations are conducted at a carrier frequency of 695 MHz, with a sampling rate of 6.144 MHz and a 10 MHz RF bandwidth. This carrier frequency is representative of operation near the upper portion of the UHF broadcast band and adjacent spectrum, and the corresponding Doppler conditions scale with carrier frequency. For the B2X mobile receiver speeds listed in Table \ref{channels}, the resulting Doppler shifts span from a few hertz in low mobility and indoor scenarios to several hundred hertz under high-speed-train (HST) scenarios. 
 \par The HST-SFN scenario is modeled using MATLAB's (nrHSTChannel) function in accordance with the high-speed reception specifications specified in 3GPP TS 38.101-4 \cite{3gpp_ts_38_101_4}, using a four-tap channel configuration and an Inter-Site Distance (ISD) of 10 km. Receiver robustness is evaluated under both ideal (without impairment) synchronization and controlled CFO and SFO impairments as specified in Table \ref{channels}, enabling reproducible assessment of normal bootstrap performance under practical mobility and synchronization conditions.
\begin{figure}[t!]
\centering
\includegraphics[width=\columnwidth]{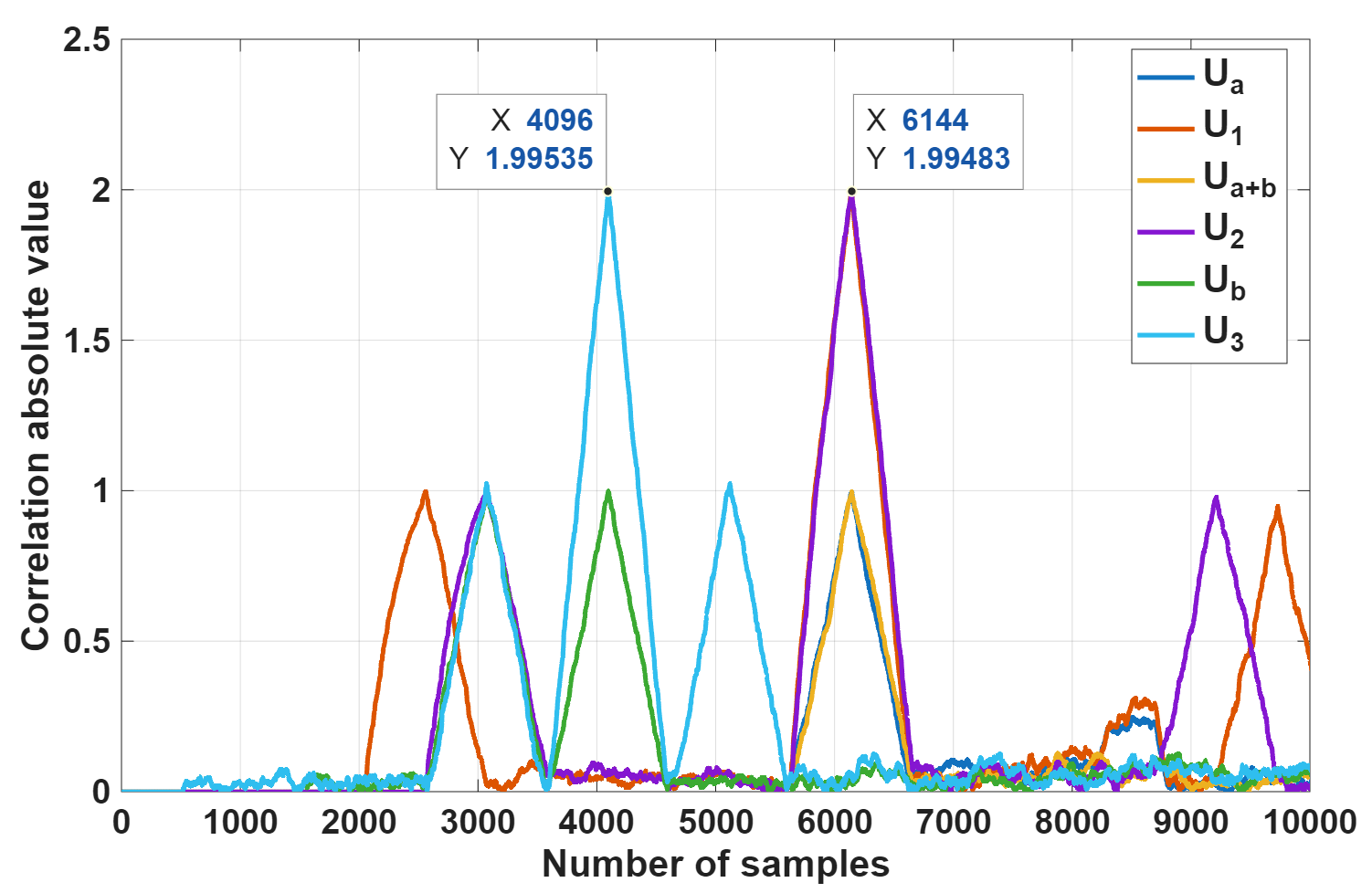}
     \vspace{-0.5cm}
      \caption{\small Peak locations of the accumulated correlation signals. 
The peaks of $U_{a}(n)$ and $U_{1}(n)$ occur at $2N_A + 2N_B + 2N_C = 6144$, 
the peaks of $U_{a+b}(n)$ and $U_{2}(n)$ occur at $2N_A + 3N_B + N_C = 6144$, 
and the peaks of $U_{b}(n)$ and $U_{3}(n)$ occur at $N_A + 3N_B + N_C = 4096$.}
      \label{add3}
\end{figure}
\begin{figure}[t!]
\centering
\includegraphics[width=\columnwidth]{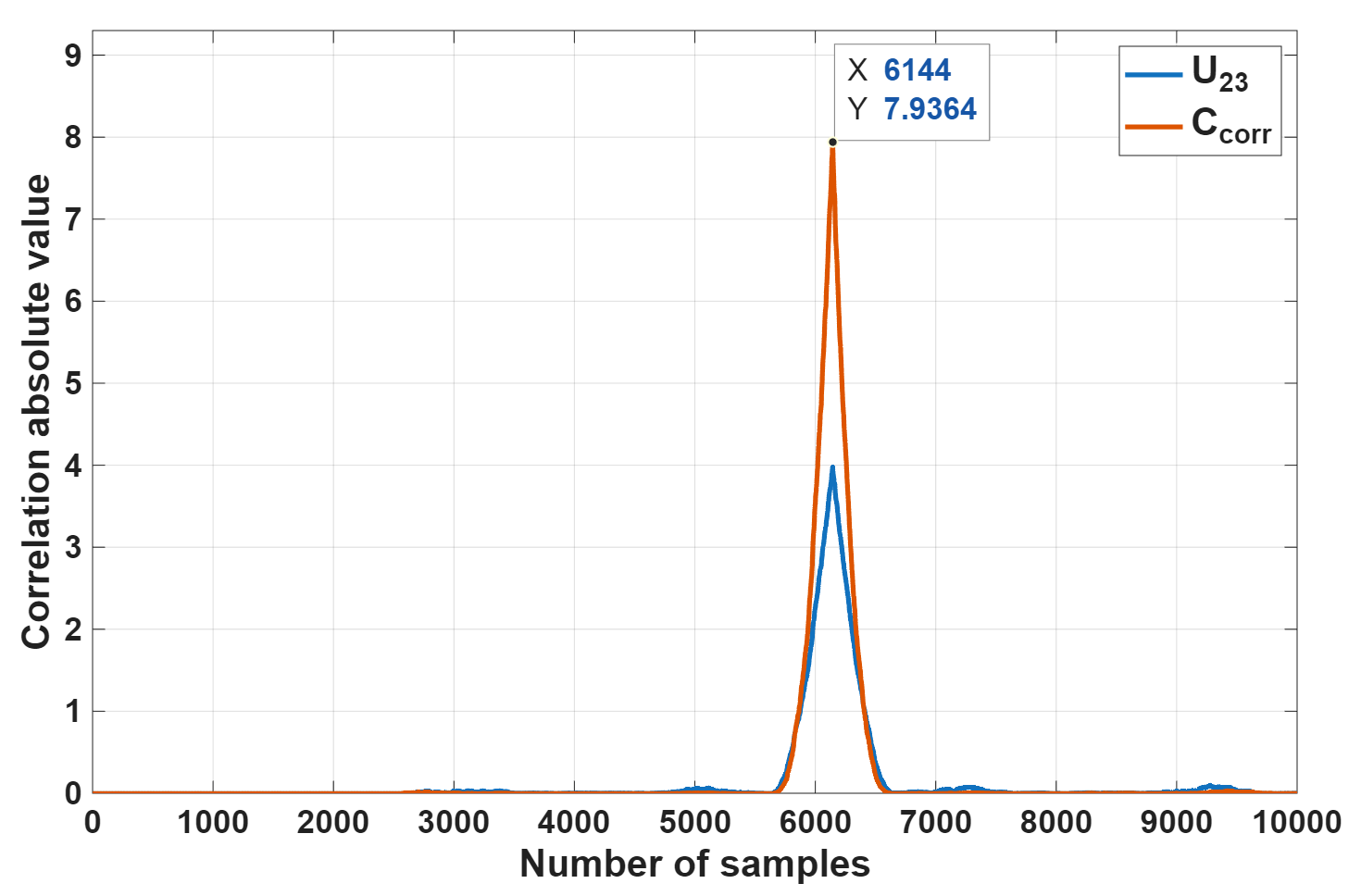}
\caption{\small Peak of the $U_{23}(n)$ signal occurs at
$2N_A+3N_B+N_C=6144$, and the peak of the
$C_{\mathrm{corr}}(n)$ signal occurs at
$2N_A+2N_B+2N_C=6144$.}
\label{finalCorr}
\end{figure}
\subsection{Results and discussion} \label{VII-C} 
{This subsection presents the simulation results (refer to Fig.~\ref{add3} and Fig.~\ref{finalCorr}) to validate the mathematical analysis of the B2X receiver given in Section \ref{IVa}} and performance evaluation results corresponding to the channel models and simulation parameters summarized in Table \ref{channels}. {The detection and decoding performance of VFS and SS is quantified in terms of the probability of frame error $P_e$ (defined in Section~\ref{VI-B_CEandDecoding})} as a function of the Signal-to-Noise Ratio (SNR), with a $P_e$  threshold of ($10^{-3}$) used to determine the required SNR. {All simulations are conducted over 100,000 frames}, with 5000 random noise samples at every frame start. Results are primarily reported for a practical receiver performing synchronization, as recommended for realistic performance assessment. For reference, selected results are reported under perfect synchronization, assuming ideal knowledge of the channel, timing, and frequency offsets at the receiver.
\begin{figure}[t!]
\centering
\vspace{-0.2cm}
\includegraphics[width=\columnwidth]{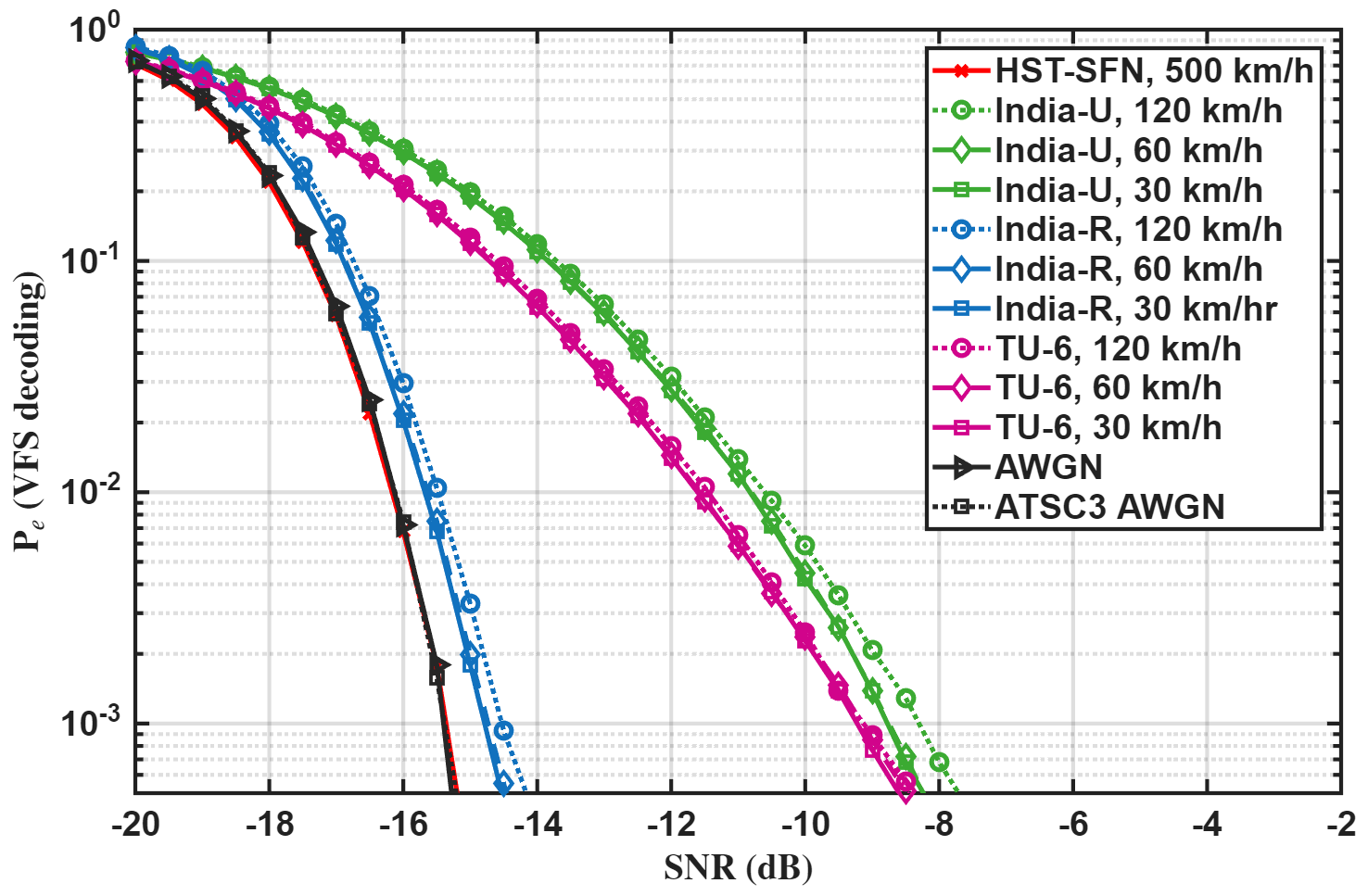}
\vspace{-0.5cm}
\caption{\small Perfect Synchronization}
\label{perfectSync_ch1}
\vspace{-0.3cm}
\end{figure}%
\begin{figure}[t!]
\centering
\vspace{-0.2cm}
\includegraphics[width=\columnwidth]{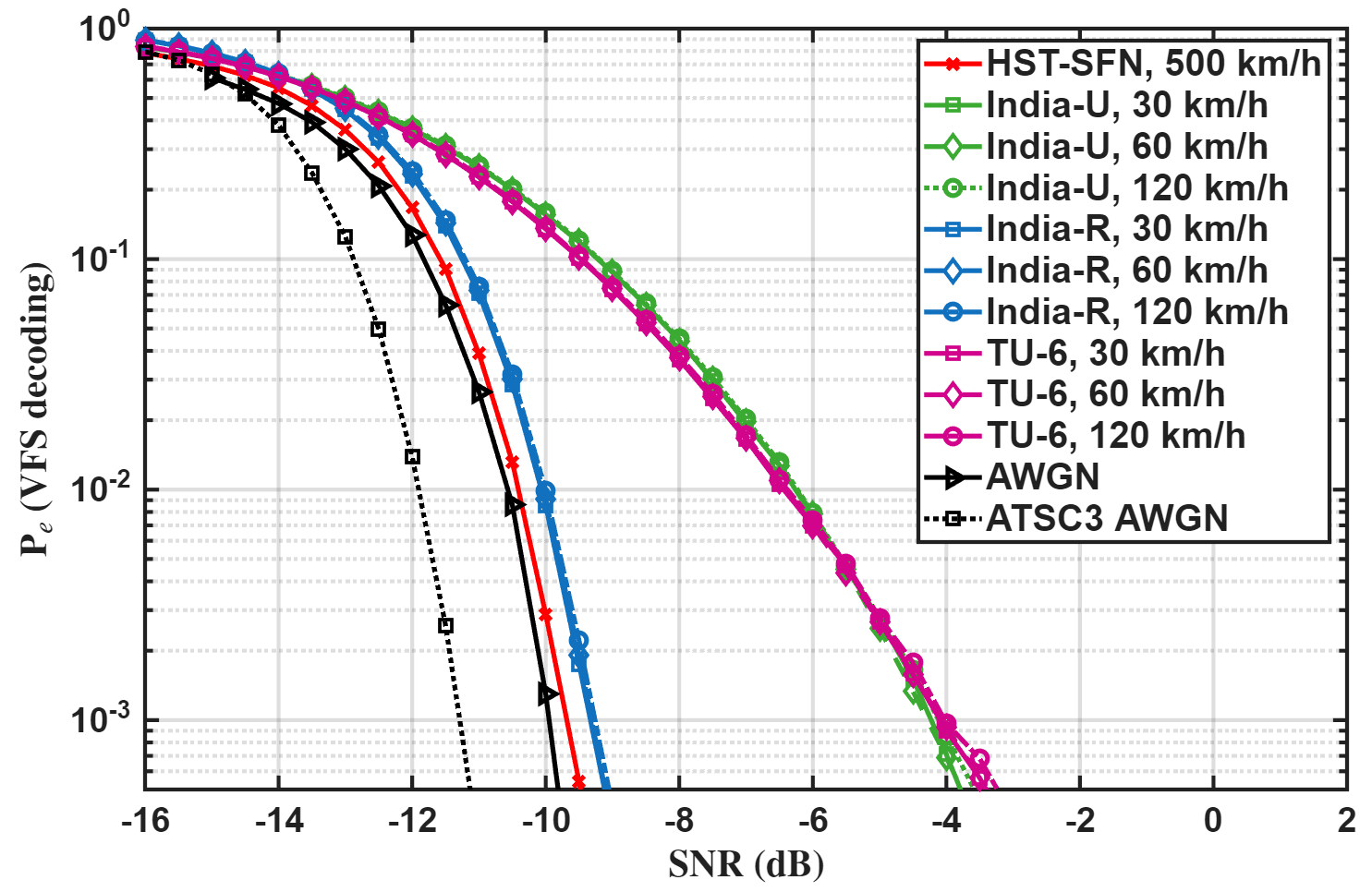}
\vspace{-0.5cm}
\caption{\small Practical Synchronization}
\label{performSync_ch1}
\vspace{-0.6cm}
\end{figure}
\par
\begin{figure}[t!]
\centering
\includegraphics[width=\columnwidth]{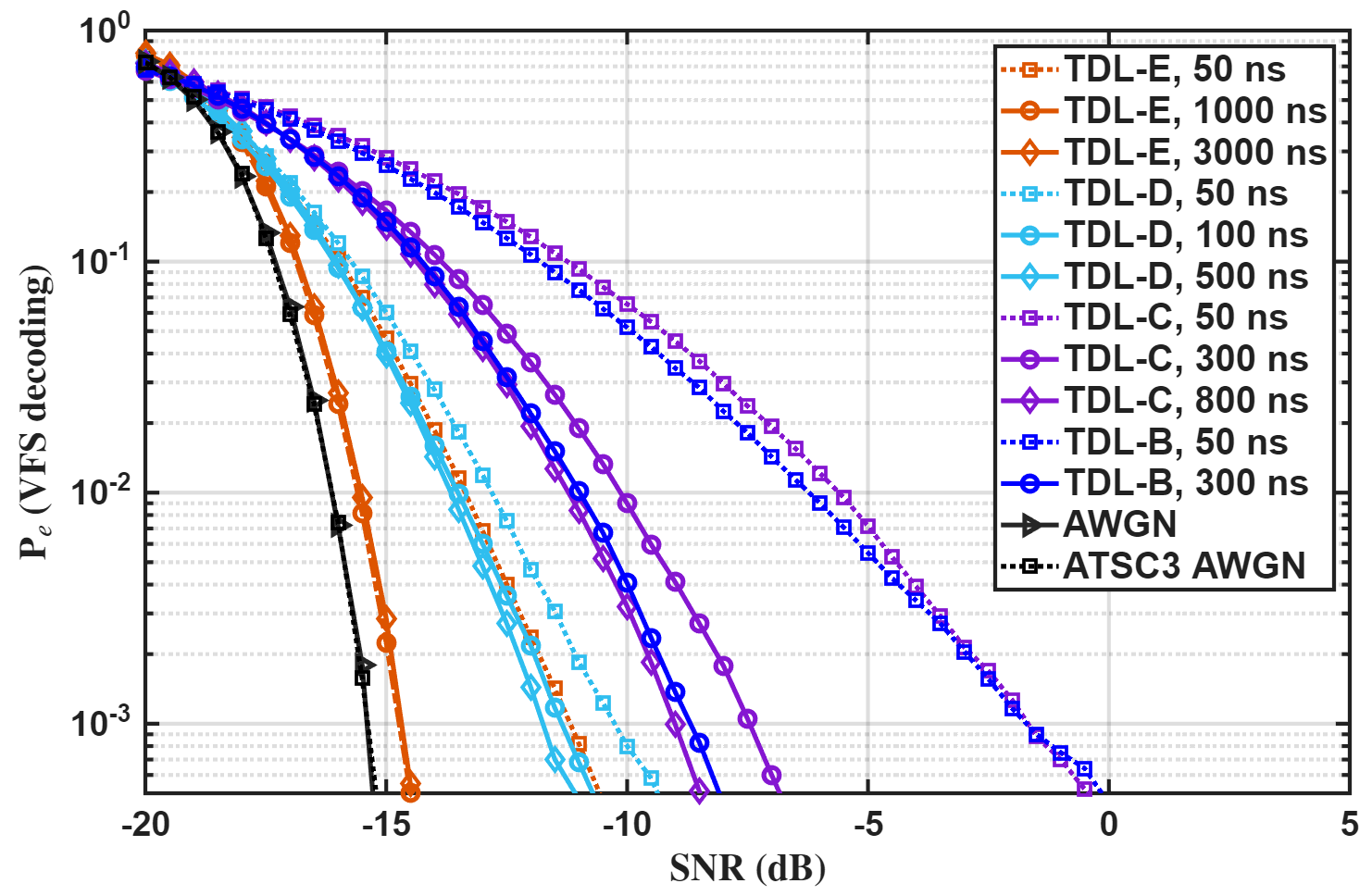}
\vspace{-0.5cm}
\caption{\small TDL Channels with Perfect Synchronization }
\label{perfectSync_tdl}
\vspace{-0.4cm}
\end{figure}

\begin{figure}[t!]
\vspace{-0.1cm}
\centering
\includegraphics[width=\columnwidth]{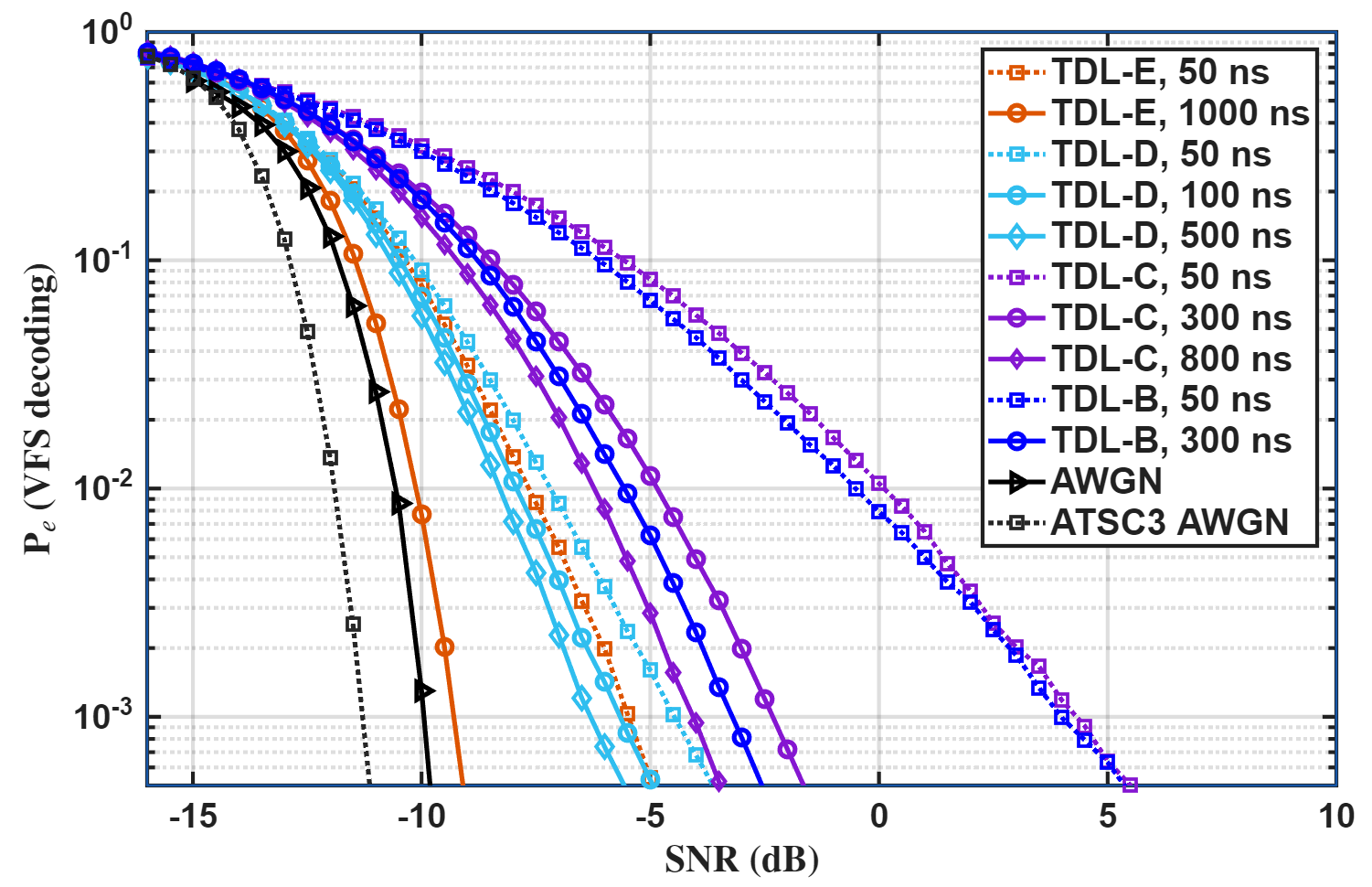}
\vspace{-0.5cm}
\caption{\small TDL Channels with Practical Synchronization}
\label{performSync_tdl}
\end{figure}

 {As a baseline, ATSC 3.0 bootstrap decoding\footnote{For a fair comparison with B2X, an ATSC 3.0 bootstrap error is declared only when an error occurs in the 8 signaling bits of the second bootstrap symbol. Errors in the third and fourth bootstrap symbols are not considered.} is evaluated under the same AWGN conditions. As shown in Figs.~\ref{perfectSync_ch1}--\ref{performSync_tdl}, both systems exhibit essentially identical performance under perfect synchronization, when only VFS decoding is evaluated. Under practical synchronization, B2X exhibits approximately 1.5 dB of additional loss when compared to ATSC 3.0, since it uses only two bootstrap symbols for synchronization as compared with four used in ATSC 3.0. Despite this reduced synchronization gain, the B2X bootstrap theoretically requires 3 dB less power, demonstrating a favorable trade-off between reduced bootstrap power and synchronization performance.}
\par Fig.~\ref{perfectSync_ch1} presents the B2X receiver (BXE) performance under perfect synchronization for TU-6, India-U, and India-R channels at speeds ranging from 30~km/h to 120~km/h to assess Doppler impact. At 120~km/h, the performance degradation is limited to approximately 0.3~dB, which further reduces to about 0.05~dB when practical synchronization is applied (see Fig.~\ref{performSync_ch1}), indicating strong Doppler robustness of the BXE. High-mobility performance is further evaluated using the HST-SFN channel at 500~km/h, where the achieved SNR ($-15.25$~dB) remains close to the AWGN baseline ($-15.38$~dB). 
\par In the practical synchronization case shown in Fig.~\ref{performSync_ch1}, a nearly constant SNR penalty of about 5~dB is observed across all channels, primarily due to residual synchronization and channel estimation errors. Nevertheless, the relative performance trends remain unchanged, with India-U and TU-6 exhibiting similar performance (around $-4$~dB) due to comparable delay and Doppler spreads, while the lower delay spread and higher channel stability of India-R result in superior VFS decoding performance ($-9.21$~dB).
\begin{figure}[t!]
\centering
\vspace{-0.2cm}
\includegraphics[width=\columnwidth]{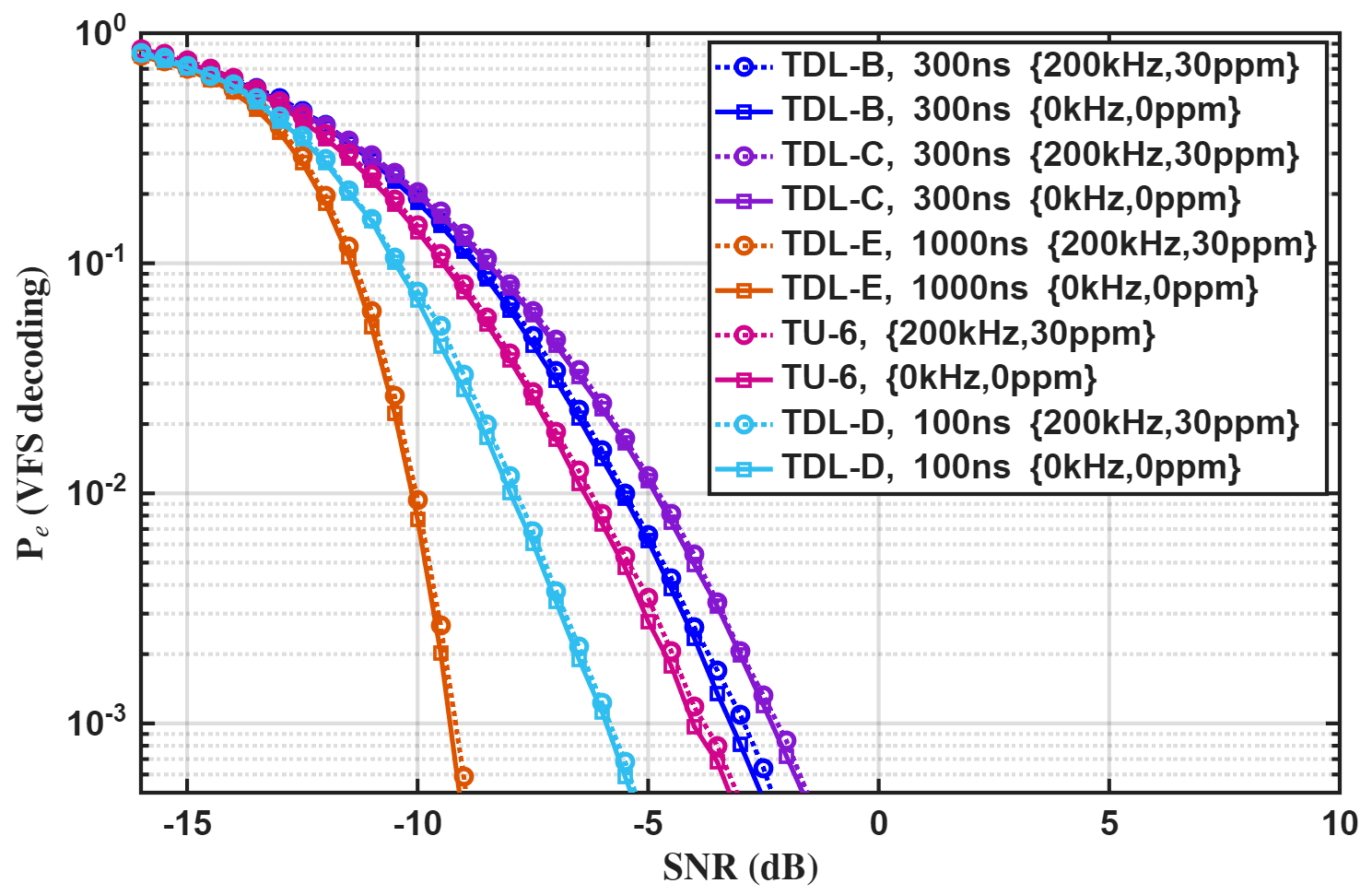}
\vspace{-0.5cm}
\caption{\small Practical Synchronization with CFO and SFO Impairments}
\label{performSync_withOffset}
\vspace{-0.3cm}
\end{figure}

\begin{figure}[t!]
\centering
\includegraphics[width=\columnwidth]{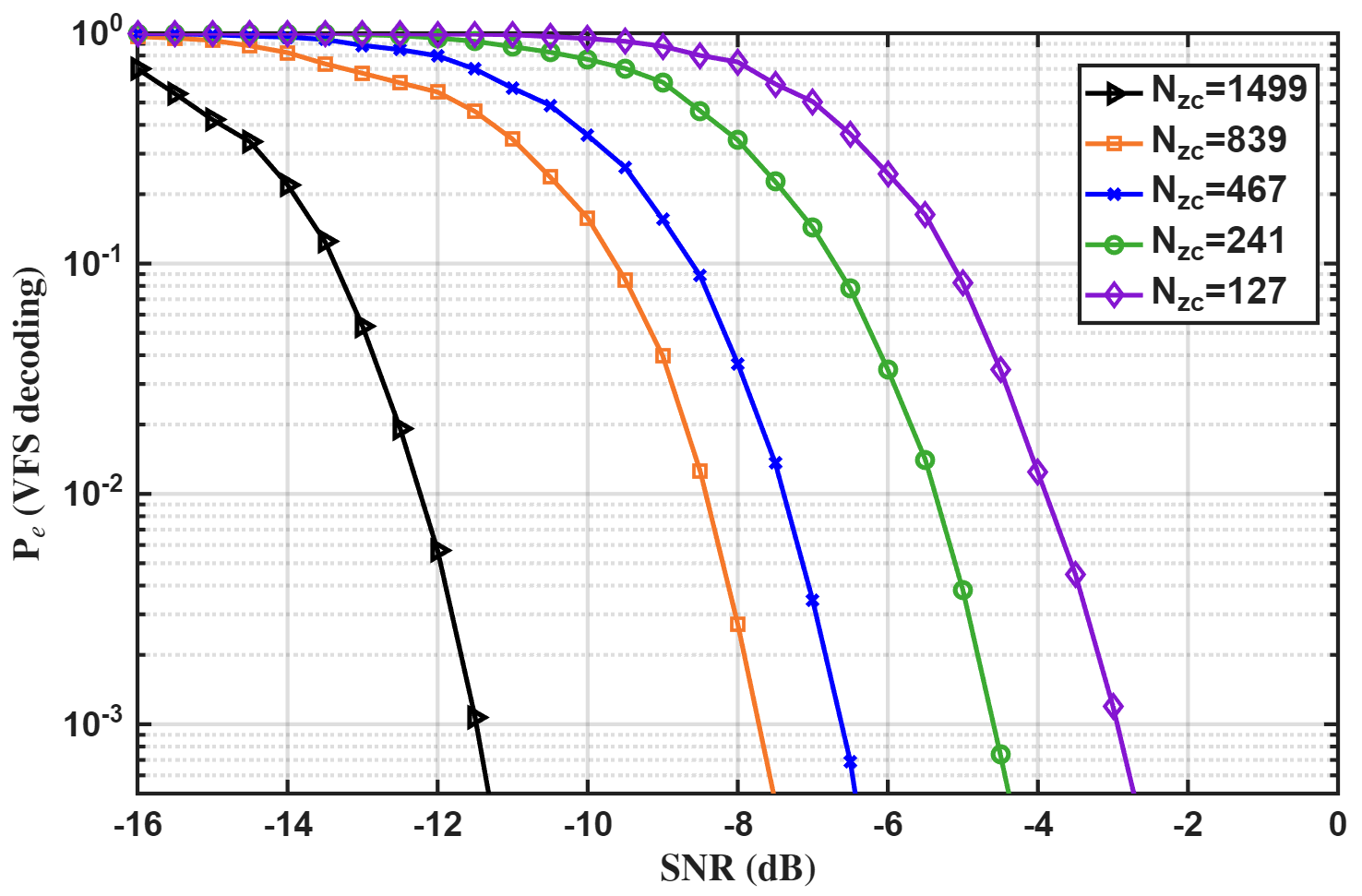}
\vspace{-0.5cm}
\caption{\small AWGN Channel, Scalable Bootstrap with Practical Synchronization}
\vspace{-0.7cm}
\label{AWGN_scalable_bootstrap}
\end{figure}
\par Figs.~\ref{perfectSync_tdl} and \ref{performSync_tdl} reveal a non-intuitive trend in VFS decoding performance across TDL channel profiles, where increasing the Root Mean Square (RMS) delay spread for a given model consistently improves decoding performance, as indicated by a leftward shift of the curves. This behaviour arises from enhanced multipath diversity in channels with larger delay spreads, where signal energy is distributed over multiple independently faded paths, enabling diversity gains at the receiver. In contrast, channels with small delay spreads (e.g., 50 ns) offer limited diversity and exhibit nearly flat fading behaviour, resulting in increased sensitivity to deep fades. Furthermore, when comparing perfect synchronization (Fig.~\ref{perfectSync_tdl}) with practical synchronization (Fig.~\ref{performSync_tdl}), all curves experience a rightward shift, indicating an SNR penalty due to synchronization errors. This penalty is more noticeable in low-delay-spread channels, where the lack of diversity makes the system more sensitive to timing inaccuracies.
\par To evaluate the impact of CFO and SFO on VFS decoding performance, simulations are carried out for TU‑6 and TDL channel models under frequency impairments, as shown in Fig.~\ref{performSync_withOffset}. The results indicate that a CFO of 200 kHz combined with an SFO of 30 ppm does not cause any noticeable degradation in decoding performance across all considered channels. The delayed correlator in Fig.~\ref{delayedCorrelation} can satisfactorily detect the CFO, which is subsequently corrected. The effect of SFO on two adjacent VFS symbols is negligible. This demonstrates that the B2X bootstrap design is inherently robust to practical frequency synchronization impairments.
Further, it is also observed that, for the normal bootstrap configuration, the SS detection performance evaluated under practical synchronization conditions is close to the VFS detection performance.
\par To evaluate the performance of the scalable bootstrap approach, we conducted additional simulations for the five bootstrap configurations ($N_{\mathrm{zc}}$\! =\! 1499, 839, 467, 241, 127) shown in Fig.~\ref{b2xScaledVFS} over an AWGN channel, as shown in Fig.~\ref{AWGN_scalable_bootstrap}. The results indicate that VFS decoding performance degrades as the bootstrap bandwidth decreases. This degradation is primarily due to the reduced signal bandwidth and FFT size (2048 to 128), which limits the number of time-domain and frequency-domain samples available for correlation-based VFS detection. The reduced number of samples limits noise averaging in the correlation-based detection process, thereby increasing the variance and producing broader, less distinct correlation peaks. Consequently, the narrower bandwidth part requires a higher SNR\footnote{The higher SNR requirement reflects the equal-power assumption used for all bootstrap configurations. In practice, additional transmit power allocated to the narrowband bootstrap can partially compensate for its reduced processing gain, making it suitable for low-complexity, energy-constrained IoT receivers. This is broadly consistent with the reduced-bandwidth synchronization mechanisms adopted for 5G NR RedCap\cite{3GPP_RedCap}.
} to achieve reliable detection, highlighting the trade-off between bandwidth scalability and detection robustness.



\section{Conclusion}
This paper presents a detailed analysis of the B2X bootstrap signal and the associated receiver processing architecture and algorithms for system discovery and synchronization. The structure of the B2X bootstrap has been examined in relation to ATSC~3.0, highlighting both continuity in detection principles and key structural and signaling differences that enable scalable operation. The analysis considers both the normal (wideband) bootstrap configuration with $N_{\mathrm{zc}}\!=\!1499$ and scalable bootstrap configurations supporting reduced receiver bandwidths. The scalable bootstrap evaluation demonstrates the trade-off between receiver bandwidth and detection robustness, while retaining a common receiver-processing framework across the supported configurations.
The results demonstrate that bootstrap detection and synchronization behavior are primarily governed by the time-domain structure of the signal, with limited sensitivity to specific ZC and PN sequence parameters. Comprehensive performance evaluation across diverse channel models, propagation conditions, and mobility scenarios confirms the robustness of the bootstrap signal and the receiver processing approach, and provides a detailed characterization of system behavior under practical deployment conditions.

The B2X bootstrap defined in the A/392-21 specification~\cite{A39221} has been realized through waveform generation consistent with the specification, validating the signal construction and transmission aspects of the design. Complementary receiver implementations have been developed to detect and decode the bootstrap signal, demonstrating that practical receivers can be constructed based on the specified waveform. The analytical, simulation, and implementation results presented in this paper provide a coherent and experimentally supported characterization of bootstrap performance, and offer a comprehensive reference for future development of B2X receivers and interpretation of the specification.

The results demonstrate robust B2X bootstrap performance within the sub-1-GHz frequency range considered in this paper. Under channel models such as TU6 and high-speed train conditions, the bootstrap and receiver processing maintain robustness in challenging propagation and mobility scenarios, supporting reliable operation beyond traditional stationary or low-mobility broadcast assumptions. {Further work will extend the scalable bootstrap evaluation to a broader range of channel and mobility conditions, particularly those representative of narrowband IoT environments, and investigate the energy-efficiency benefits of scalable bootstrap configurations for low-power devices. }

\section*{Acknowledgment}
The authors acknowledge the support of Sinclair and EdgeBeam Wireless for funding our work. We thank Joe Fabiano, Mark Aitken, Sangsu Kim, Sesh Simha and Prof. Prasanna Chaporkar for their engagement and contributions to the broader B2X development effort.

\ifCLASSOPTIONcaptionsoff
  \newpage
\fi



%
\bibliographystyle{IEEEtran}
\bibliography{convergence.bib}

@ARTICLE{ibc,
author="{Michael Simon, Sangsu Kim, James Kelso and Joe Fabiano}",
journal={IBC},
title="{ATSC (B2X) Multicast Broadcast Neutral-Host O-RAN System Architecture}",
year={2025},
volume={},
number={},
pages={},
doi={}}

@INPROCEEDINGS{kamran2022,
author={Kamran, Rashmi and Jha, Pranav and Kiran, Shwetha and Karandikar, Abhay and Chaporkar, Prasanna and Saha, Anindya and Chakraborty, Arindam},
booktitle={2022 National Conference on Communications (NCC)},
title="{A Survey on Multicast Broadcast Services in 5G and Beyond}",
year={2022},
volume={},
number={},
pages={344-349},
doi={10.1109/NCC55593.2022.9806729}}

@misc{streamable,
  title        = {{Survey: Almost 90\% of Gen Z Viewers Watch Video on Phones Weekly}},
  author       = {{The Streamable}},
  year         = {2023},
  howpublished = {{\url{http://thestreamable.com/survey-almost-90-percent-of-gen-z-viewers-watch-video-on-phones-weekly-are-mobile-streaming-plans-coming-to-us}}},
  note         = {Accessed: 2026-03-28}
}

@misc{muvi,
  title        = {Streaming Statistics and Growth Projections for 2026},
  author       = {Sreejata Basu},
  year         = {2025},
  howpublished = {\url{https://www.muvi.com/blogs/streaming-statistics-in-2025/}},
  note         = {Published: November 26, 2025. Accessed: 2026-03-28}
}

@ARTICLE{ericcson,
author={Ericsson},
journal={Report},
title="{Ericsson Mobility Report}",
year={June, 2025},
volume={},
number={},
pages={},
doi={}}

@article{standard2016system,
  title={System discovery and signaling},
  author={ATSC},
  journal={A/321},
  month={March},
  year={2016}
}

@article{Chernock_2016,
 title={{ATSC 3.0 Next Generation Digital TV Standard—An Overview and Preview of the Issue}},
 volume={62},
 ISSN={1557-9611},
 url={},
 DOI={10.1109/tbc.2016.2515542},
 number={1},
 journal={IEEE Transactions on Broadcasting},
 publisher={Institute of Electrical and Electronics Engineers (IEEE)},
 author={Chernock,
 Rich and Gomez-Barquero,
 David and Whitaker,
 Jerry and Park,
 Sung-Ik and Wu,
 Yiyan},
 year={2016},
 month=mar,
 pages={154–158} }

@article{Fay_2016,
 title={{An Overview of the ATSC 3.0 Physical Layer Specification}},
 volume={62},
 ISSN={1557-9611},
 url={},
 DOI={10.1109/tbc.2015.2505417},
 number={1},
 journal={IEEE Transactions on Broadcasting},
 publisher={Institute of Electrical and Electronics Engineers (IEEE)},
 author={Fay,
 Luke and Michael,
 Lachlan and Gomez-Barquero,
 David and Ammar,
 Nejib and Caldwell,
 M. Winston},
 year={2016}, month=mar, pages={159–171} }

@article{Earnshaw_2016,
 title={{Physical Layer Framing for ATSC 3.0}},
 volume={62},
 ISSN={1557-9611},
 url={},
 DOI={10.1109/tbc.2016.2518621},
 number={1},
 journal={IEEE Transactions on Broadcasting},
 publisher={Institute of Electrical and Electronics Engineers (IEEE)},
 author={Earnshaw,
 Mark and Shelby,
 Kevin and Lee,
 Hakju and Oh, Youngho and Simon, Michael}, year={2016}, month=mar, pages={263–270} }

@article{He_2016,
 title={{System Discovery and Signaling Transmission Using Bootstrap in ATSC 3.0}},
 volume={62},
 ISSN={1557-9611},
 url={},
 DOI={10.1109/tbc.2016.2518619},
 number={1},
 journal={IEEE Transactions on Broadcasting},
 publisher={Institute of Electrical and Electronics Engineers (IEEE)}, author={He, Dazhi and Shelby, Kevin and Earnshaw, Mark and Huang, Yihang and Xu, Honglian and Park, Sung-Ik}, year={2016}, month=mar, pages={172–180} }

@article{Kim_2019,
 title={{A Novel Iterative Detection Scheme of Bootstrap Signals for ATSC 3.0 System}},
 volume={65},
 ISSN={1557-9611},
 url={},
 DOI={10.1109/tbc.2018.2855660},
 number={2}, journal={IEEE Transactions on Broadcasting}, publisher={Institute of Electrical and Electronics Engineers (IEEE)}, author={{H. Kim, J. Kim, S. -I. Park, N. Hur, M. Simon and M. Aitken}}, year={2019}, month=jun, pages={211–219} }

@article{Kwon_2025,
 title={{Comparative Assessment of Physical Layer Performance: ATSC 3.0 vs. 5G Broadcast in Laboratory and Field Tests}},
 volume={71},
 ISSN={1557-9611},
 url={},
 DOI={10.1109/tbc.2024.3482183},
 number={1}, journal={IEEE Transactions on Broadcasting}, publisher={Institute of Electrical and Electronics Engineers (IEEE)}, author={{S. Kwon et al.}}, year={2025}, month=mar, pages={2–10} }

@article{Ahn_2023,
 title={{Evaluation of ATSC 3.0 and 3GPP Rel-17 5G Broadcasting Systems for Mobile Handheld Applications}},
 volume={69},
 ISSN={1557-9611},
 url={},
 DOI={10.1109/tbc.2022.3222988},
 number={2}, journal={IEEE Transactions on Broadcasting}, publisher={Institute of Electrical and Electronics Engineers (IEEE)}, author={{S. -K. Ahn et al.}}, year={2023}, month=jun, pages={338–356} }

@techreport{etsi2004102,
  author       = {{ETSI TS 102 831 V1.2.1}},
  title        = {{Digital Video Broadcasting (DVB); Implementation Guidelines for a Second Generation Digital Terrestrial Television Broadcasting System (DVB-T2)}},
  institution  = {},
  number       = {},
  year         = {2012},
  month        = {Aug}
}

@article{ahn2022characterization,
  title={{Characterization and modeling of UHF wireless channel in terrestrial SFN environments: Urban fading profiles}},
  author={{S. Ahn et al.}},
  journal={IEEE Transactions on Broadcasting},
  volume={68},
  number={4},
  pages={803--818},
  year={2022},
  publisher={IEEE}
}

@techreport{3gpp_tr38901_v18,
  author       = {{3GPP TR 38.901 V18.0.0}},
  title        = {{Study on Channel Model for Frequencies from 0.5 to 100 GHz (Release 18)}},
  institution  = {},
  type         = {Technical Report},
  number       = {},
  year         = {2024},
  month        = mar,
}

@techreport{3gpp_ts_38_101_4,
 author       = {{3GPP TS 38.101-4}},
  title        = {{NR; User Equipment (UE) radio transmission and reception; Part 4: Performance requirements}},
  institution  = {},
  type         = {Technical Specification},
  number       = {},
  year         = {2022},
 
}

@techreport{cost207_1989,
  author       = {{COST 207 Management Committee}},
  title        = {{COST 207: Digital Land Mobile Radio Communications -- Final Report}},
  institution  = {{Commission of the European Communities}},
  address      = {Brussels, Belgium},
  year         = {1989},
  pages        = {135--147},
  note         = {Channel models for digital land mobile radio communications (including TU-6)}
}

@misc{A39221,
  author       = {{ATSC}},
  title        = {{ATSC Candidate Standard: B2X System Discovery and Signaling}},
  howpublished = {Doc. S44-2-002r7},
  year         = {2026},
  month        = {July},
  day          = {9},
  note         = {\url{https://www.atsc.org/wp-content/uploads/2026/07/S44-2-002r7-A392-21-2026-05-14_System_Discovery_and_Signalling.pdf}, Accessed: Aug. 14, 2026}
}

@techreport{IITB_whitePaper,
  author      = {Raj Kumar Thenua and Pranay Jha and Rashmi Kamran and Kumar Appaiah and Prasanna Chaporkar},
  title       = {{Reimagining Broadcasting in India with ATSC 3.0 B2X: Use Cases, Requirements and Opportunities}},
  institution = {IIT Bombay},
  year        = {2025},
  month       = dec,
  type        = {White Paper}
}

@INPROCEEDINGS{R4Ref1,
  author={Kamran, Rashmi and Kelso, James and Jha, Pranav and Kiran, Shwetha and Fabiano, Leonard J and Chaporkar, Prasanna and Kim, Sangsu and Simha, Sesh},
  booktitle={2025 IEEE Future Networks World Forum (FNWF)}, 
  title={{Convergence of 6G and ATSC3 Networks via Broadcast to Everything (B2X): Use cases, Solutions and Collaborations}}, 
  year={2025},
  volume={},
  number={},
  pages={1-6},
  doi={10.1109/FNWF66845.2025.11317205}}

@INPROCEEDINGS{R4Ref2,
  author={Sahu, Harsh and Sharma, Manas and K, Priyadarsini and Shah, Preksha and Kamran, Rashmi and Fabiano, Leonard J and Kim, Sangsu},
  booktitle={2026 35th Wireless and Optical Communications Conference (WOCC)}, 
  title={{Intent-Driven AI-Native Network Slicing for Rural Broadcasting over ATSC 3.0 / B2X: A Reinforcement Learning Approach}}, 
  year={2026},
  volume={},
  number={},
  pages={1-5},
  doi={10.1109/WOCC69802.2026.11556186}}

@misc{3GPP_RedCap,
  author       = {{3GPP}},
  title        = {{A Glimpse into RedCap NR devices}},
  howpublished = {\url{https://www.3gpp.org/technologies/nr-redcap-glimpse}},
  note         = {Accessed: Aug. 19, 2026}
}
\end{document}